\documentclass[traditabstract,times]{aa} 

\newcommand{\ergps}{erg\thinspace s$^{-1}$}
\newcommand{\phpspsqcm}{ph\thinspace cm$^{-2}$\thinspace s$^{-1}$}
\newcommand{\ergpspsqcm}{erg\thinspace s$^{-1}$\thinspace cm$^{-2}$}
\newcommand{\psqcm}{cm$^{-2}$}
\newcommand{\nH}{$N_{\rm H}$}

\usepackage{graphicx}
\begin{document}
\title{The XMM deep survey in the CDF-S VIII. X-ray properties of the two brightest sources}


   \author{K. Iwasawa\inst{1}\thanks{Email: kazushi.iwasawa@icc.ub.edu}
          \and
C. Vignali\inst{2,3}
\and
A. Comastri\inst{2}
\and
R. Gilli\inst{2}
\and
F. Vito\inst{2,3}
\and 
W.N. Brandt\inst{4}
\and
F.J. Carrera\inst{5}
\and
G. Lanzuisi\inst{2}
\and
S. Falocco\inst{6,7}
\and
F. Vagnetti\inst{8}
}

\institute{ICREA and Institut de Ci\`encies del Cosmos, Universitat de Barcelona, IEEC-UB, Mart\'i i Franqu\`es, 1, 08028 Barcelona, Spain
         \and
INAF - Osservatorio Astronomico di Bologna, Via Ranzani, 1, 40127 Bologna, Italy
\and
Universit\`a di Bologna - Dipartimento di Astronomia, Via Ranzani, 1, 40127 Bologna, Italy
\and
Department of Astronomy and Astrophysics, The Pennsylvania State University, 525 Davey Lab, University Park, PA 16802, USA
\and
Instituto de F\'isica de Cantabria (CSIC-UC), Avenida de los Castros, 39005, Santander, Spain
\and
University Federico II, Via Cintia, Building 6, 80126, Napoli, Italy
\and
INFN Napoli, Via Cintia, 80126, Napoli, Italy
\and
Dipartimento di Fisica, Universit\`a di Roma “Tor Vergata”, Via della Ricerca Scientifica 1, 00133 Roma, Italy
          }


 
          \abstract{We present results from the deep XMM-Newton
            observations of the two brightest X-ray sources in the
            Chandra Deep Field South (CDFS), PID 203 ($z=0.544$) and
            PID 319 ($z=0.742$). The long exposure of 2.5 Ms over a 10
            year period (net 4 yr with a 6 yr gap) makes it possible
            to obtain high quality X-ray spectra of these two Type I
            AGN with X-ray luminosity of $10^{44}$ \ergps, typical
            luminosity for low-redshift PG quasars, track their X-ray
            variability both in flux and spectral shape. Both sources
            showed X-ray flux variability of $\sim $10-20 \% in rms
            which is similar in the soft (0.5-2 keV) and hard (2-7
            keV) bands. PID 203, which has evidence for optical
            extinction, shows modest amount of absorption (\nH$\leq
            1\times 10^{21}$ \psqcm) in the X-ray spectrum. Fe K
            emission is strongly detected in both objects with EW$\sim
            0.2$ keV. The lines in both objects are moderately broad
            and exhibit marginal evidence for variability in shape and
            flux, indicating that the bulk of the line emission come
            from their accretion disks rather than distant tori.}

\keywords{X-rays: galaxies - Galaxies: active - Surveys
                             }
\titlerunning{Two bright X-ray sources in XMM-CDFS}
\authorrunning{K. Iwasawa et al.}
   \maketitle
%

\section{Introduction}

X-ray spectra of active galaxies provide us with a probe of the
physical conditions of the innermost part of the active
nucleus. However, sufficient details of their X-ray spectra, e.g.,
absorption, Fe K emission, can be obtained only for nearby, bright
objects with the limited sensitivity available from the X-ray
telescopes currently in operation. Possible exceptions are extremely
luminous QSOs and objects in survey fields with very long
exposures. The deep XMM-Newton observations of the Chandra Deep Field
South (CDFS, Giacconi et al 2002) with a 2.5 Ms exposure (Comastri et
al 2011) give us one of the exceptional opportunities to access to
active galactic nuclei (AGN) with a moderate QSO luminosity beyond the
nearby Universe. Here, we present the XMM-Newton data on the two
brightest X-ray sources ($f_{\rm X}\sim 10^{-13}$ \ergpspsqcm) in the
XMM-CDFS (Ranalli et al 2013), PID 203 (z=0.544) and PID 319 (z=0.742)
with the X-ray luminosities of $\sim 10^{44}$ \ergps\ (Table 1).

These two sources are exceptionally bright among the X-ray sources in
the field. In terms of source counts in the full XMM-Newton band (e.g,
0.4-10 keV) accumulated over the whole exposures, the third brightest
source\footnote{The third brightest source is PID 358 in the 2-10 keV
  source catalogue by Ranalli et al (2013). Note that the counts
  listed in the catalogue are corrected for the exposure map (i.e.,
  telescope vignetting etc.), not those observed counts referred
  here. PID 358 is located at the edge of the field where the
  correction is large while the two sources in this work are located
  in the central part of the field. The absorbed spectrum of PID 358
  (Comastri et al, in prep.) amplifies the difference in counts from
  the two unabsorbed sources presented in this article when the energy
  range lower thatn 2 keV is considered.} falls an order of magnitude
below the two. In order to place the two X-ray sources in context
among the general AGN samples, PID 203 and PID 319 along with sources
in the XMM-CDFS (Comastri et al, in prep.), PG quasars (Piconcelli et
al 2005), SDSS quasars (Strateva et al 2005), the CAIXA AGN
(duplicated PG quasars in Piconcelli et al 2005 were removed, Bianchi
et al 2009) are plotted on the $L_{\rm X}$-$z$ plane in Fig. 1. The
CAIXA sample consists of bright AGN in the XMM-Newton archive, and
together with the PG quasars, they represent X-ray sources with
sufficient counts to warrant a resonably constraining spectral
abalysis. The SDSS quasars are another flux-limited sample of X-ray
AGN. It is clear from Fig. 1 that PID 203 and PID 319 stand out the
XMM-CDFS sample, and have the typical X-ray luminosity of low
redshifts ($z=$0.1-0.2) PG quasars but are located at higher
redhsifts. The two sources lie in the upper envelope of the SDSS
quasars, yet a detailed spectral study is still possible.


\begin{figure}
\centerline{\includegraphics[width=0.4\textwidth,angle=0]{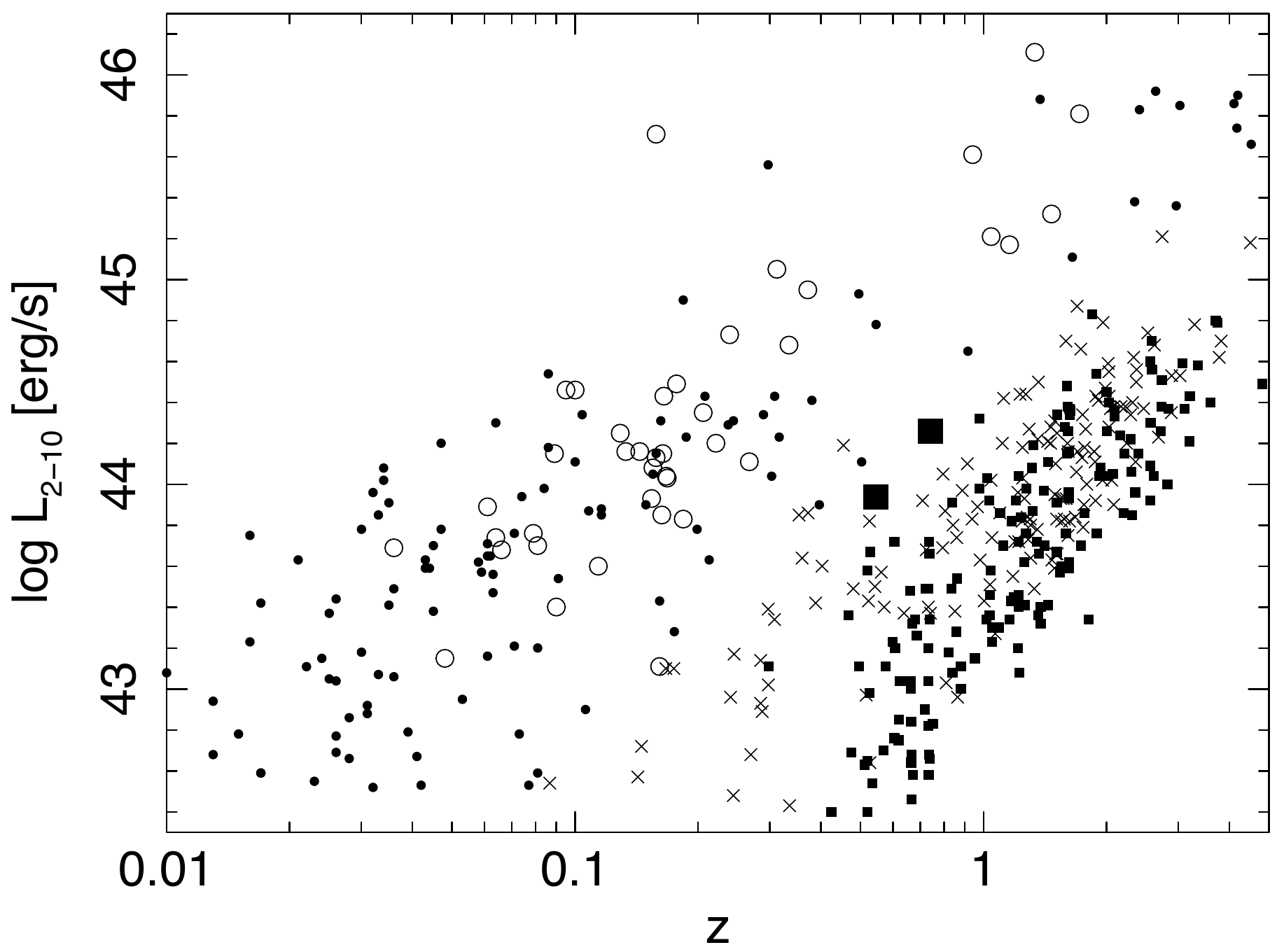}}
\caption{PID 203 and PID 319 (large filled squares) are plotted along
  with various AGN samples on the $L_{\rm X}$-$z$ plane: XMM-CDFS
  (filled squares, Comastri et al in prep., Ranalli et al 2013), PG
  quasars (open circles, Piconcelli et al 2005), SDSS quasars
  (crosses, Strateva et al 2005), and CAIXA (filled circles, Bianchi
  et al 2009). The $L_{\rm X}$ is the rest-frame 2-10 keV
  luminosity. The monochromatic 2 keV luminosities in Strateva et al
  (2005) were converted to 2-10 keV luminosities assuming a power-law
  spectrum of $\Gamma = 1.8$. }
\end{figure}

Although the quality of the data does not rival that for nearby,
bright AGN, it gives more than a rough spectral shape and its
long-term variability, including that of their Fe K lines. The
previous study of AGN Fe K line properties beyond the local Universe,
in particular, resorted to a stacking analysis (e.g.,
Streblyanska et al 2005; Corral et al 2008, Iwasawa et al 2012;
Chaudhary et al 2012; Falocco et al. 2012). These XMM-Newton data of
the two CDFS sources offer reasonable quality of Fe K line data for an
individual AGN at $z=$0.5-0.7.
The 4~Ms data of the CDFS obtained with the Chandra X-ray Observatory
(Xue et al. 2011) were added for investigating the X-ray flux and Fe K
line variability.

The cosmology adopted here is $H_0=70$ km s$^{-1}$ Mpc$^{-1}$,
$\Omega_{\Lambda}=0.72$, $\Omega_{\rm M}=0.28$.

\section{Observations}

The CDFS was observed with XMM-Newton with 33 exposures in the years of
2001-2002 and 2008-2010. Full details of the observations are found in
Ranalli et al (2013). We used the data taken from the three EPIC
cameras (pn, MOS1 and MOS2) after the standard filtering of background
flaring intervals. The net exposures, integrated counts, mean fluxes
and luminosities for the two sources are given in Table 2.

Due to the visibility of the field, observations were performed in either
summer (July-August) or winter (January-February) periods. In order to
obtain sufficient quality for a spectral study, the 33 exposures were
grouped into six XMM-intervals (X1-X6 in Table 3), which give a half-yr
sampling of time-scale in the 2008-2010 period.  

The two bright X-ray sources in the field have the source
identification numbers PID 203 and PID 319 in the XMM-CDFS source
catalogue (Ranalli et al 2013), for which basic information and the
Chandra counterparts are given in Table 1.

The data from the three EPIC cameras were used jointly in the analysis
presented below unless stated otherwise. In the pn camera, the sources
fall in the chip gap in a few observations, which resulted in a
shorter exposure time than the full exposure of the field. PID 203 was
always observed with the two MOS cameras. After 2009, one CCD chip of
the MOS1 camera lost sensitivity. PID 319 fell on that CCD chip
during a few observations, reducing the exposure time. The background
fraction of the total (net source + background) counts accumulated for
both sources during the observations is 8-9 \% in the 0.5-7 keV
band. The stability of the energy scale during the XMM-Newton
exposures has been verified to be within 10 eV at the energy of 8 keV,
using the instrumental background line of Cu (M. Guainazzi,
priv. comm.).

The Chandra 4 Ms data of the CDFS (Xue et al 2011) were combined when
investigating flux variability. Since the XMM-Newton and Chandra
spectra still have inter-calibration problems, for example, in spectral
slope (e.g., Lanzuisi et al. 2013), the Chandra data were used only
for flux variability. For the moment, the Chandra observations of CDFS
span over the period of 2000-2010 and are divided into four intervals
(C1-C4 in Table 3).

\begin{table}
\begin{center}
\caption{Basic properties of the two XMM-CDFS sources.}
\begin{tabular}{ccccc}
PID & $z$ & RA & Dec. & Other ID \\
 & & deg. & deg. & \\[5pt]
203 & 0.544$^a$ & 53.03594 & $-27.79297$ & CDF-63$^a$, XID\thinspace 141$^c$ \\
319 & 0.742$^b$ & 53.11233 & $-27.68497$ & CDF-42a$^a$, XID\thinspace 375$^c$ \\
\end{tabular}
\begin{list}{}{}
\item[] $^a$ Szokoly et al 2004; $^b$ Silverman et al 2010; $^c$ Xue et al 2011.
\end{list}
\end{center}
\end{table}

\begin{table}
\begin{center}
\caption{The XMM-Newton observations}
\begin{tabular}{cccccc}
PID & Exposure & Counts & $f_{\rm 0.5-2}$ & $f_{\rm 2-7}$ & $L_{\rm 2-10}$ \\
& (1) & (2) & (3) & (4) & (5) \\[5pt]
203 & 1.86/2.31/2.84 & 50/21/22 & 5.5 & 5.6 & 0.87 \\
319 & 1.74/2.49/2.30 & 45/17/16 & 6.3 & 6.0 & 1.8 \\
\end{tabular}
\begin{list}{}{}
\item[] Note --- (1) Exposure time in $10^6$ s for the EPIC
  pn/MOS1/MOS2 cameras; (2) Integrated net counts in units of $10^3$
  ct in the 0.5-7 keV band collected by the pn/MOS1/MOS2 cameras; (3)
  0.5-2 keV flux in unit of $10^{-14}$ \ergpspsqcm, as observed; (4)
  2-7 keV flux in unit of $10^{-14}$ \ergpspsqcm, as observed; and
  (5) Rest-frame 2-10 keV luminosity corrected for absorption in unit
  of $10^{44}$ \ergps.
\end{list} 
\end{center}
\end{table}

\begin{table}
\begin{center}
\caption{The six XMM-Newton and four Chandra observation intervals of CDFS.}
\begin{tabular}{clc}
X/C & Period & Exposure \\
(1) & (2) & (3) \\[5pt]
C1 & 2000 May-Jun & 0.19 \\
C2 & 2000 Dec & 0.64 \\
X1 & 2001 Jul & 0.07 \\
X2 & 2002 Jan & 0.36 \\
C3 & 2007 Sep-Nov & 0.97 \\
X3 & 2008 Jul & 0.41 \\
X4 & 2009 Jan & 0.68 \\
X5 & 2009 Jul & 0.42 \\
X6 & 2010 Jan-Feb & 0.80 \\
C4 & 2010 Mar-Jul & 1.96 \\
\end{tabular}
\begin{list}{}{}
\item[] Note --- (1) XMM-Newton (X) or Chandra (C) observation
  intervals; (2) Period of the observation interval; (2) Exposure time
  in $10^6$ s (the maximum value among the three EPIC cameras for the
  XMM-Newton intervals).
\end{list}
\end{center}
\end{table}

\section{PID 203}



\begin{figure}
\centerline{\includegraphics[width=0.4\textwidth,angle=0]{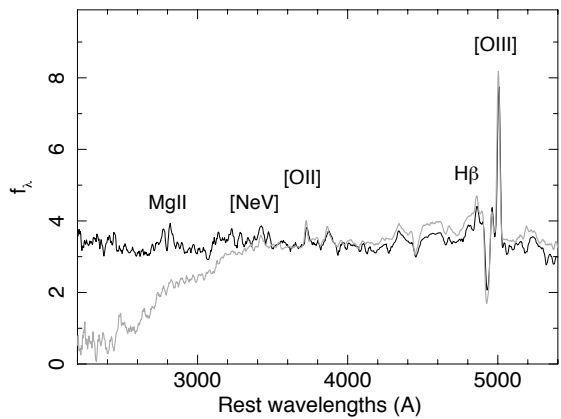}}
\caption{The UV/optical spectra of PID203, taken at the VLT with the
  FORS spectrographs (Szokoly et al 2004). The horizontal axis is the
  rest-frame wavelengths in \AA, assuming the galaxy redshift
  of $z=0.544$. The flux density $f_{\lambda }$ is in units of
  $10^{-17}$ erg cm$^{-2}$ s$^{-1}$ \AA$^{-1}$. The data were taken at
  two occasions, 2000 Oct 28 (black line) and 2000 Nov 23 (grey
  line). The spectrum became redder in the second observation than in
  the first observation.}
\end{figure}

PID 203 is the second brightest X-ray source in the XMM-CDFS field
after PID 319 in terms of flux but a larger number of counts were
collected as it was more centrally located on the detector than PID
319 during the pointings.  The X-ray source is classified optically as
broad-line AGN ($z=0.544$) in Szokoly et al (2004) with a detection of
broad Mg {\sc ii} (Fig. 2). However, the optical spectrum shows
self-absorbing features in the emission-lines (Mg {\sc ii} and H$\beta
$) and a redder continuum than optically-selected QSOs normally
exhibit, indicating the presence of moderate extinction. The two
spectra presented in Szokoly et al (2004) were taken on two occasions
separated by 3 weeks in time (see Fig. 2). The spectrum taken in the
second occasion shows an increased continuum redenning and deeper
self-absorption in the major lines than in the first observaion,
suggesting the optical extinction in this QSO is variable on a
relatively short time-scale. The optical extinction in the first, less
extincted spectrum is estimated to be $A_{\rm V}\simeq 2.0$, when the
dust attenuation law for the Milky Way (Cardelli, Clayton \& Mathis
1989) is adopted. The other, redder spectrum cannot be described well
by just increasing the amount of reddening but at least $A_{\rm V}\geq
4$ would be needed. 

The UV emission of PID 203 measured with the OM instrument (Antonucci
et al 2014) is found to have the mean luminosity-density of
$L_{\nu}\simeq 1.7\times 10^{29}$ erg~s$^{-1}$~Hz$^{-1}$ at 2500~\AA,
after corrction for the contribution of the host galaxy. The
optical-to-X-ray spectral slope is $\alpha_{\rm OX}\simeq -1.25\pm
0.04$ and does not change significantly between observations.


\begin{figure}
\hbox{{\includegraphics[width=0.24\textwidth,angle=0]{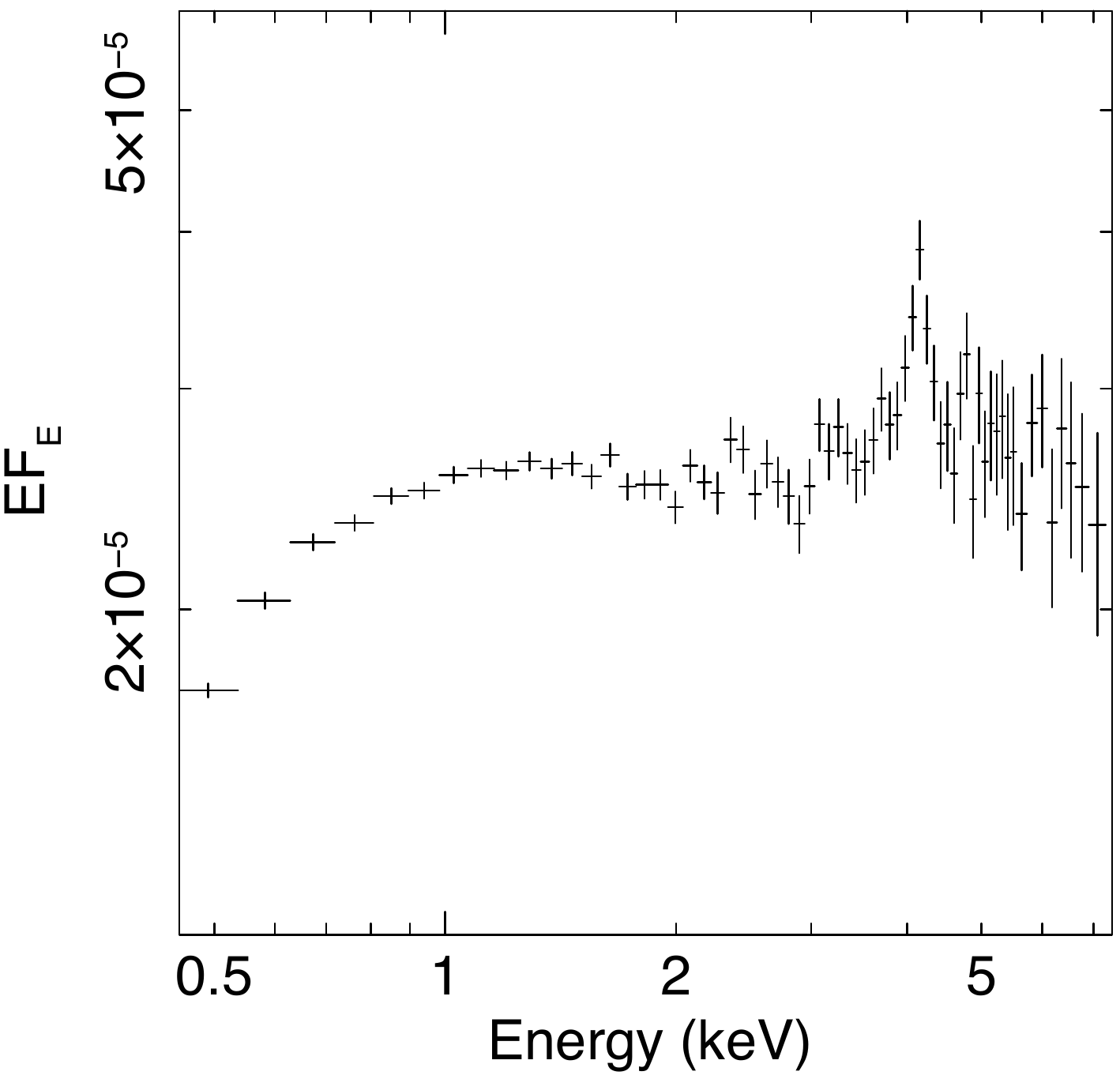}}
{\includegraphics[width=0.24\textwidth,angle=0]{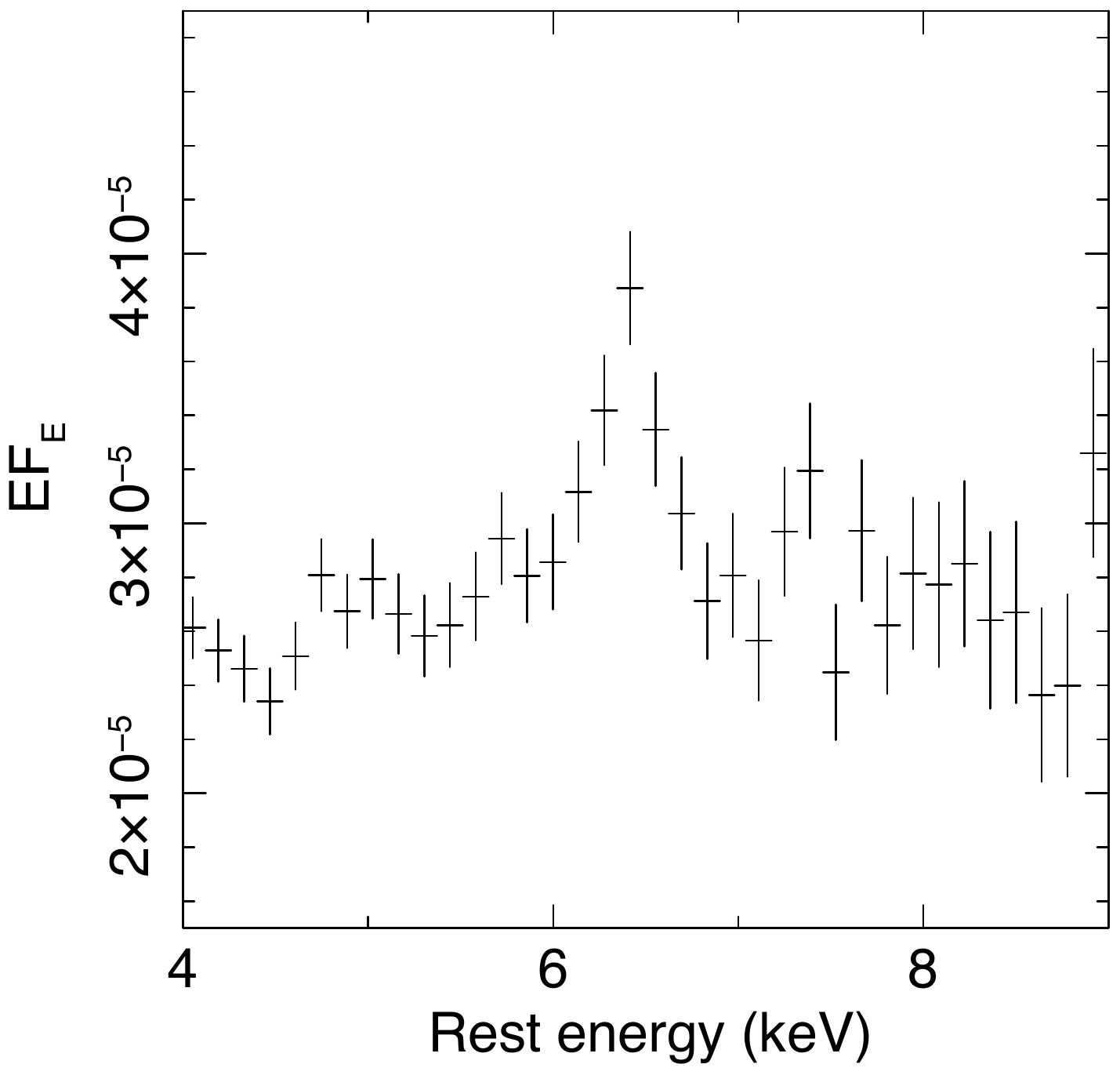}}}
\caption{Left: The 0.5-7 keV spectrum of PID 203, obtained from the
  full exposure of three EPIC cameras. The data are plotted in flux
  units of $EF_{E}$ as a function of energy as observed. The energy
  range corresponds to the rest-frame 0.77-10.8 keV. Right: Details of
  the Fe K band spectrum plotted in the rest-frame 4-9 keV range. }
\end{figure}

\subsection{The mean spectrum}

The mean X-ray spectrum integrated over the full exposure (Fig. 3)
exhibits a typical continuum slope for QSOs with photon index
$\Gamma\sim 2$, moderate amount of absorption in excess above the very
small Galactic extinction of the CDF-S ($N_{\rm H, Gal}=7\times
10^{19}$ \psqcm, Kalberla et al. 2005), and an iron K-shell (Fe K)
emission line. Fitting an absorbed power-law to the 0.5-7 keV band
data gives photon index $\Gamma = 2.00\pm 0.01$ and column density of
the cold, excess absorption measured in the galaxy rest-frame of
$N_{\rm H}=(5.5\pm 0.7)\times 10^{20}$ \psqcm. The spectral parameters
were obtained by fitting the data from the three EPIC cameras jointly
and the quoted errors are of $1 \sigma $ hereafter unless stated
otherwise.

While a single power-law agrees with the observed continuum well (see
Table 4), the relatively strong Fe K line, as detailed below, requires
a significant amount of reflection component. Including a reflection
component (modelled by {\tt reflionx} of Ross \& Fabian 2005) in the
fit, the continuum slope of the illuminating power-law was found to
have $\Gamma = 2.05\pm 0.01$ with the reflecting matter being cold
(with the ionization parameter $\xi=10$ erg cm s$^{-1}$, the lowest
value of the model). With this slightly steeper slope than that from a
single power-law fit, the cold absorption is now found to be $N_{\rm
  H}=(9.4\pm 0.8)\times 10^{20}$ \psqcm. The reflection component
carries $\sim 9$ \% of the 2-7 keV band flux, in agreement with
the albedo of cold matter covering a solid angle of $\sim 2\pi$. The
fitted amount of reflection is mainly driven by the Fe K line
strength, not by the continuum shape which would deviate from a single
power-law. Expected continuum steepening towards lower energies may be
masked by the absorption. The other source, PID 319, shows no evidence
for absorption and has a similarly strong Fe K line. We will
investigate the reflection component in more detail on the PID 319
spectrum in absence of the uncertainty due to absorption (see Section
4.1).

The measured absorbing column densities corresponds to optical
reddening of $A_{\rm V}=0.27$ or 0.47, depending on modelling, when
the Galactic $N_{\rm H}$-$A_{\rm V}$ conversion factor \nH $\simeq
2\times 10^{21}A_{\rm V}$ \psqcm (e.g., Gorenstein 1975; Predehl \&
Schmitt 1995) is adopted. Although we note that the optical and X-ray
observations were not simultaneous, the X-ray inferred nuclear
extinction is smaller than that suggested by the optical spectra. It
is contrary to what is often found in luminous AGN (e.g., Merloni et
al 2014).


The Fe K emission is clearly detected (Fig. 3).  Fitting a Gaussian
gives a line centroid of $6.39\pm 0.05$ keV in the galaxy rest-frame
with a moderate broadening of $\sigma =0.18^{+0.08}_{-0.05}$ keV
(Table 5). The rest-frame equivalent width (EW) of the line is found
to be $0.18\pm 0.03$ keV, corresponding to the line flux $(2.9\pm
0.5)\times 10^{-7}$ \phpspsqcm, derived from the Gaussian fit, is
relatively large for QSOs and in agreement with the substantial
reflection inferred from the full-band spectrum above. The good
detection of this level of line flux in AGN is only possible with the
long exposure of the XMM-CDFS.

\begin{table}
\begin{center}
\caption{The 0.5-7 keV continuum spectra of PID 203 and PID 319.}
\begin{tabular}{cccccccccc}
PID & $\Gamma $ & $N_{\rm H}$ & $\chi^2$/dof \\
& (1) & (2) & (3) \\[5pt]
203 & $2.00\pm 0.01$ & $5.5\pm 0.7$ & 461.7/502 \\
319 & $2.05\pm 0.01$ & $<0.2$ & 554.7/454 \\
\end{tabular}
\begin{list}{}{}
\item[] Note --- (1) Photon index of power-law; (2) Absorbing column
  density of cold matter in excess of the Galactic absorption ($N_{\rm
    H} = 7\times 10^{19}$ \psqcm), measured in the galaxy rest-frame
  in units of $10^{20}$ \psqcm. The upper limit for PID 319 is the
  90 \% confidence limit; (5) $\chi^2$ value and the degrees of
  freedom (dof). This $\chi^2$ was obtained when a Gaussian is
  included in the fit to describe the Fe K feature, details of which
  is shown in Table 5. This single absorbed power-law gives a good fit
  for the full band continuum of PID 203 while the poor quality of fit
  for PID 319, as indicated by the $\chi^2$ value, is caused by the
  spectral break (or curvature, see text).
\end{list}
\end{center}
\end{table}

\begin{table}
\begin{center}
\caption{The Fe K features in PID 203 and PID 319}
\begin{tabular}{ccccc}
PID & $E$ & $\sigma $ & $I$ & {\sl EW} \\
& (1) & (2) & (3) & (4) \\[5pt]
203 & $6.39\pm 0.04$ & $0.18^{+0.08}_{-0.05}$ & $2.9\pm 0.5$ & $0.18\pm 0.03$ \\
319 & $6.48\pm 0.07$ & $0.29^{+0.12}_{-0.09}$ & $4.2\pm 1.0$ & $0.20\pm 0.05$ \\
\end{tabular}
\begin{list}{}{}
\item[] Note --- The 2-6 keV data were fitted by a power-law continuum
  with a Gaussian for the Fe K line. (1) Rest-frame centroid
  energy in keV; (2) Gaussian dispersion ($\sigma $) for line width in
  keV; (3) Line intensity in units of $10^{-7}$ \phpspsqcm; (4)
  Equivalent width in keV as measured in the galaxy rest-frame.
\end{list}
\end{center}
\end{table}


\begin{figure}
\hbox{{\includegraphics[width=0.212\textwidth,angle=0]{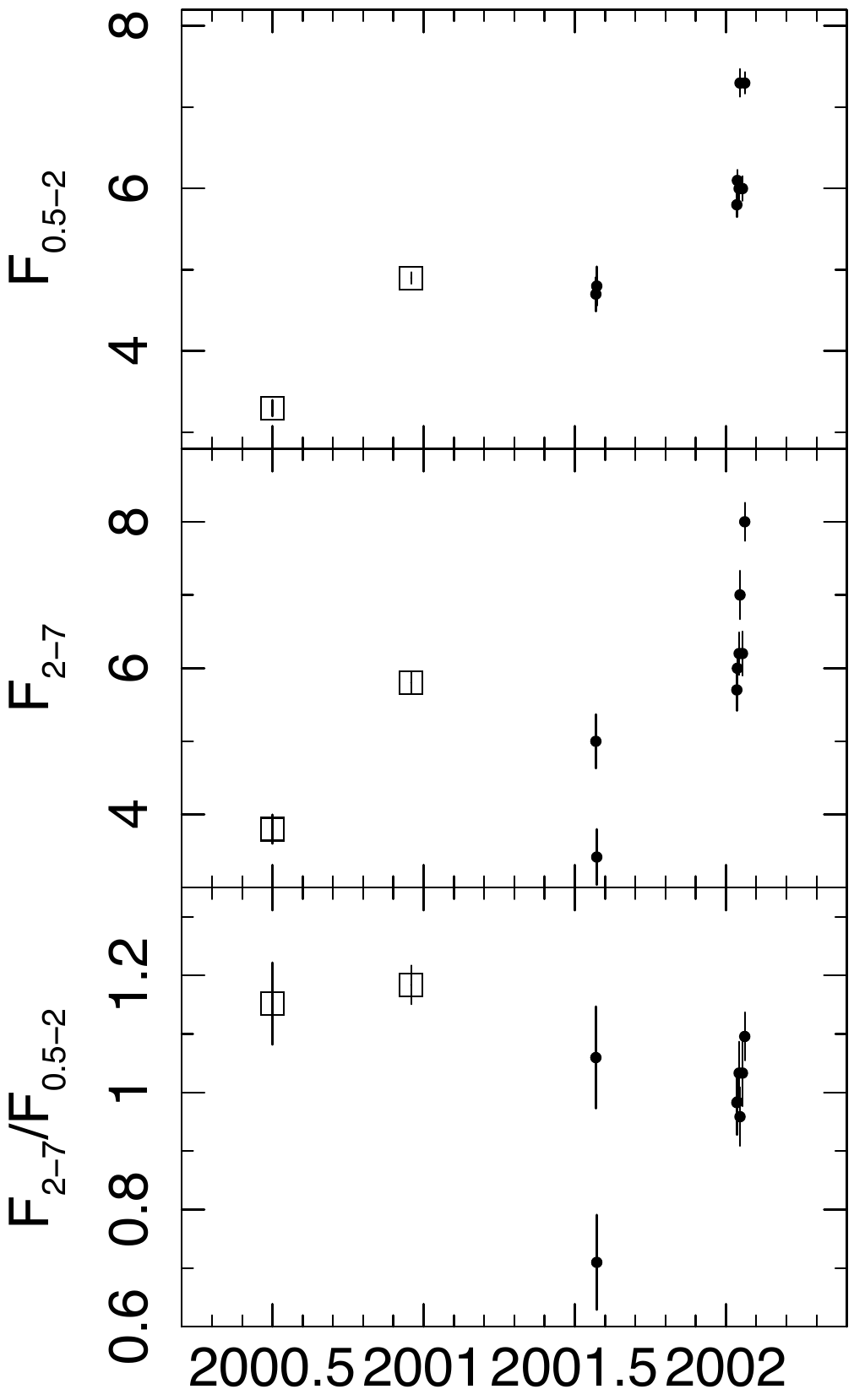}}{\includegraphics[width=0.255\textwidth,angle=0]{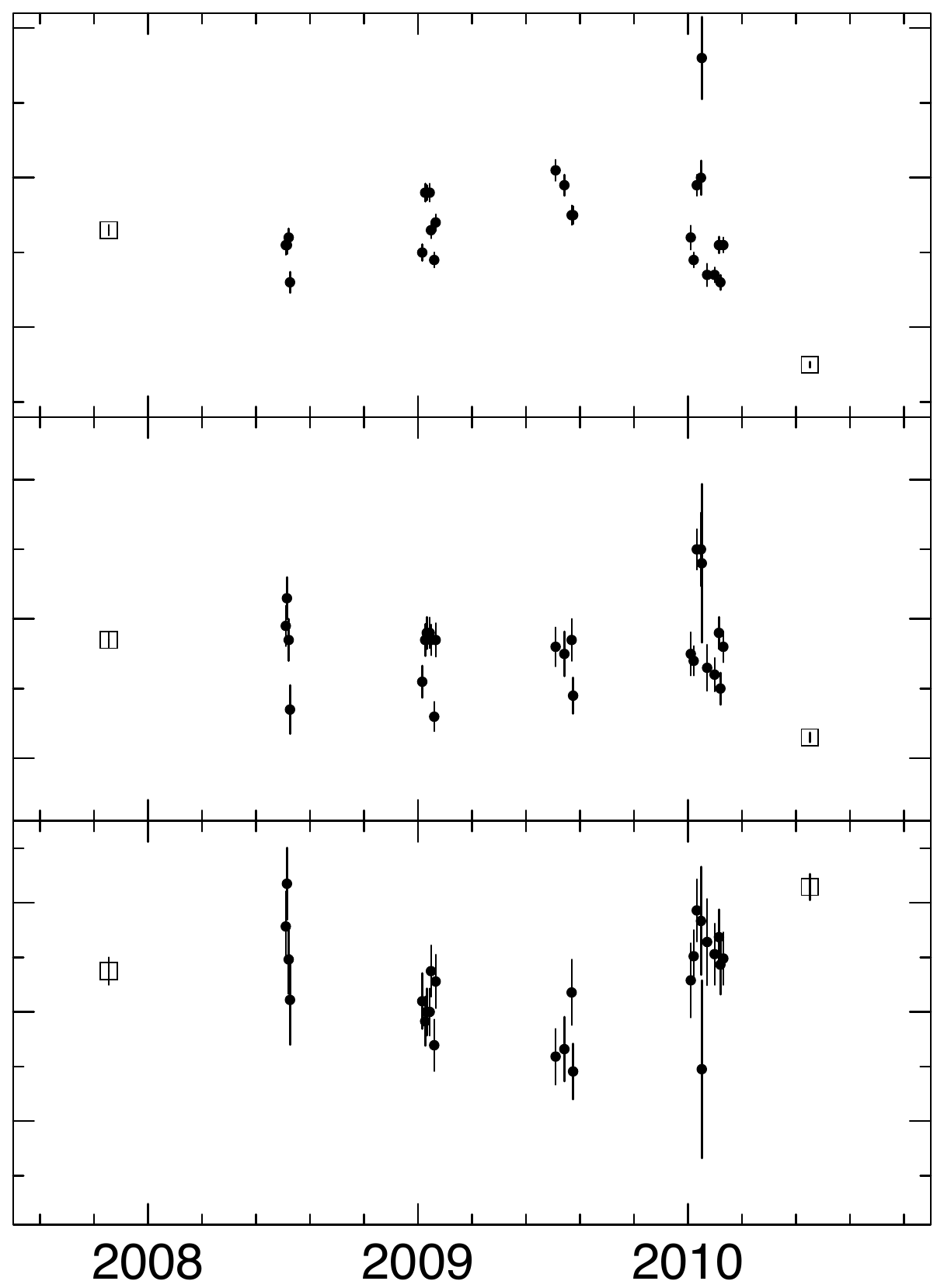}}}
\caption{The observed 0.5-2 keV and 2-7 keV flux and the flux ratio (2-7 keV/ 0.5-2 keV) variations of PID 203, obtained from the XMM-Newton (filled circles) and Chandra (open squares) observations. The time axis is in year. }
\end{figure}

\begin{figure}
\hbox{{\includegraphics[width=0.24\textwidth,angle=0]{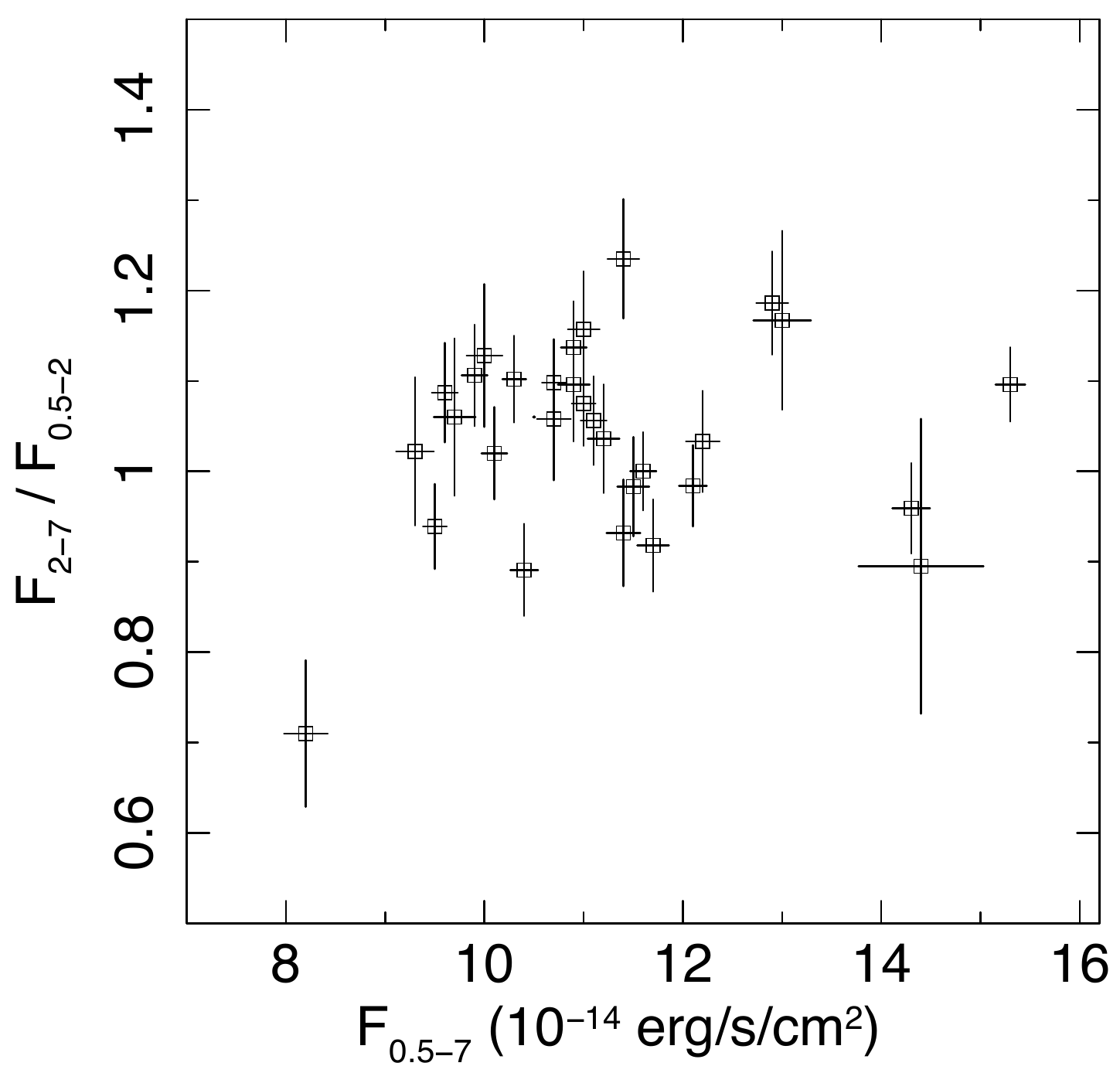}}\hspace{1mm}{\includegraphics[width=0.24\textwidth,angle=0]{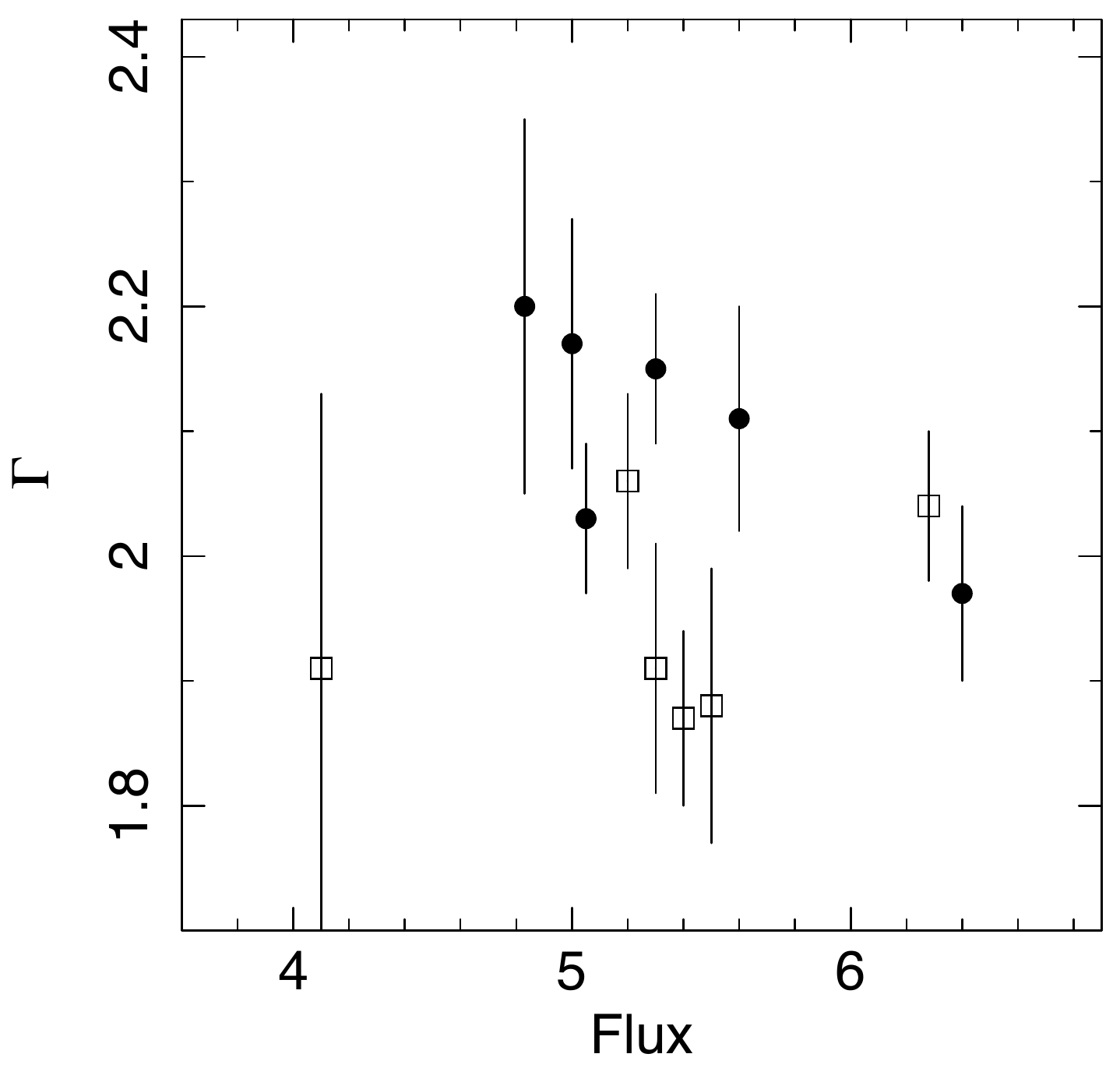}}}
\caption{Left: Plot of X-ray colour ($F_{\rm 2-7}/F_{\rm 0.5-2}$)
  against 0.5-7 keV band flux for the 33 XMM exposures for PID
  203. Right: Plot of the 0.5-2 keV (filled circles) and 2-7 keV (open
  squares) continuum slopes (photon index $\Gamma$) against the 0.5-2
  keV and 2-7 keV fluxes, respectively, in units of $10^{-14}$
  \ergpspsqcm\ for the six XMM intervals. Although a constant
  hypothesis cannot be ruled out for both slopes, the ``steeper when
  brighter'' trend often observed for AGN slope variability is not
  seen.}
\end{figure}


\begin{figure}
\centerline{\includegraphics[width=0.37\textwidth,angle=0]{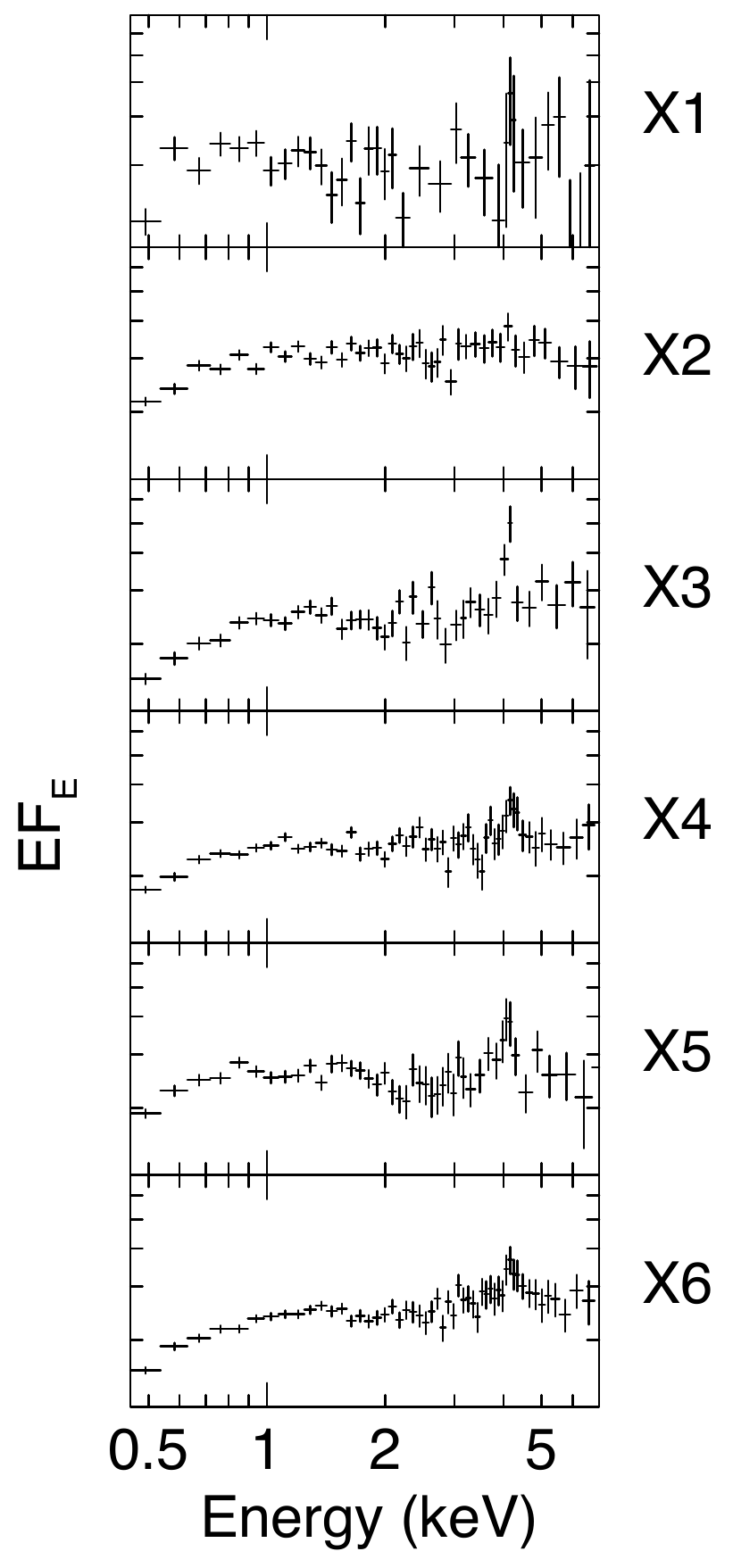}}
\caption{The 0.5-7 keV band spectra of PID 203 obtained from the three
  EPIC cameras, observed in the six XMM time-intervals (Table
  3). These spectra are plotted in the identical flux range. Possible
  Fe K line and other spectral variabilities over the six intervals can
  be visually inspected.}
\end{figure}


\begin{figure}
\centerline{\includegraphics[width=0.37\textwidth,angle=0]{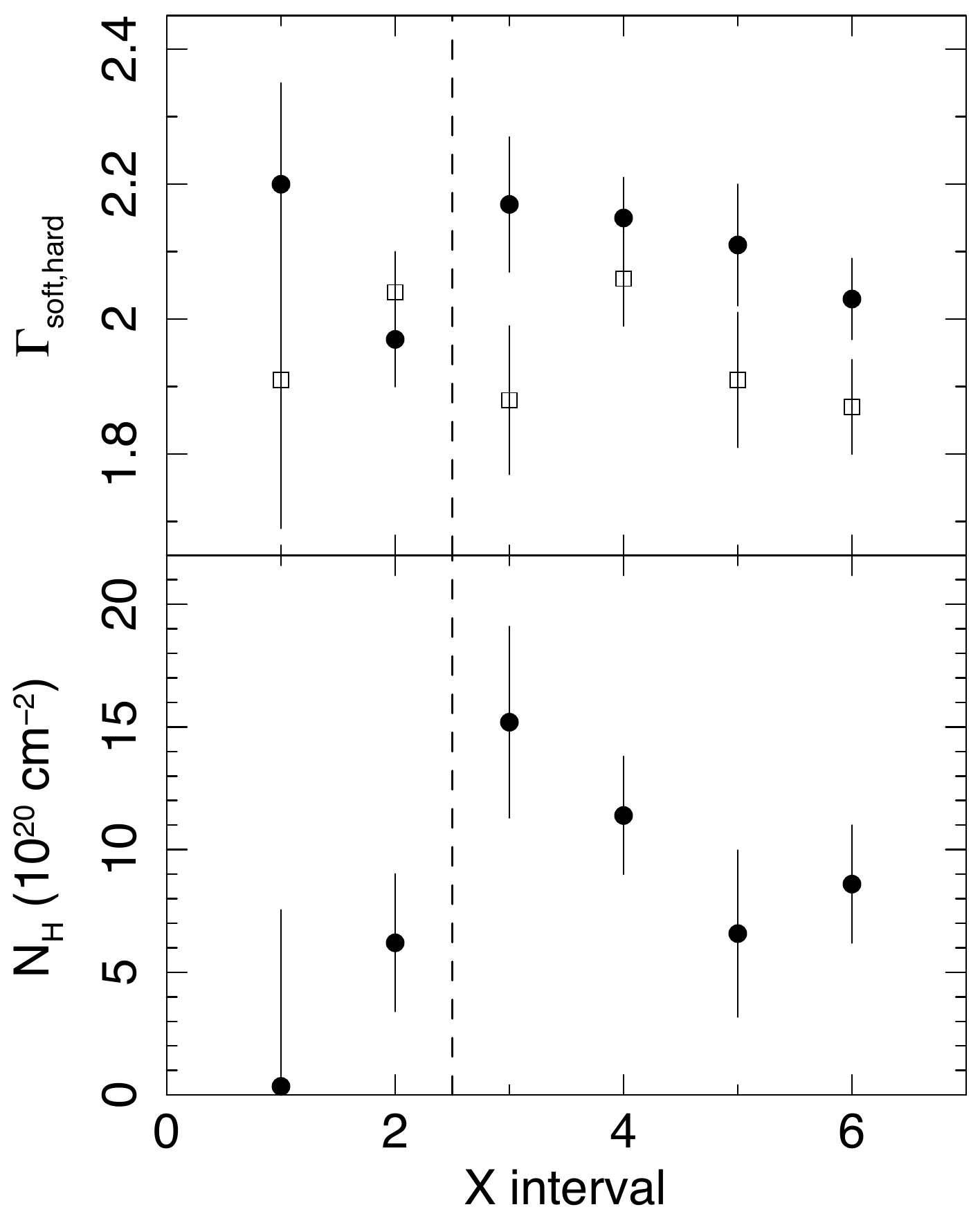}}
\caption{Upper panel: Photon indices measured in the 0.5-2 keV (filled
  circles) and 2-7 keV (open squares) bands for the six intervals of
  the XMM-Newton observations of PID 203 (X1--X6 in Table 3). The
  soft-band slopes were obtained by fitting an absorbed power-law
  while the hard band slopes by a simples power-law with a Gaussian
  for Fe K (Fig. 8). The vertical dashed-line marks the 6-yr gap
  between thr X2 and X3 intervals (see Table 3). Lower panel: The
  absorbing column density ($N_{\rm H}$) in excess of the Galactic
  value measured in the galaxy rest-frame for the six intervals. These
  $N_{\rm H}$ values were obtained by using the 0.5-2 keV data. Given
  the error bars, evidence for variability of $N_{\rm H}$ is weak with
  $\chi^2 = 6.5$ for 5 degrees of freedom.}
\end{figure}

\subsection{X-ray flux variability}

The flux of the X-ray source during each exposure was obtained by
fitting a power-law to the data. The flux measurements for all 33
XMM-Newton exposures in the observed 0.5-2 keV and 2-7 keV bands and
for the four intervals of the Chandra observations (C1-C4 in Table 3)
are plotted, along with the flux ratio $F_{2-7}/F_{0.5-2}$ in
Fig.~4. These flux measurements were without any absorption
correction. 

Taking the average within the six XMM intervals (X1-X6), ten
measurements, combined with the four Chandra data points (C1-C4),
spans over 10 yr with a roughly half-yr sampling (with a 6-yr gap in
the middle). These long-term measurements give fractional variability
amplitude (e.g., Vaughan et al. 2003), $F_{\rm var}=17\pm 5$ \%. The
25 XMM-Newton exposures of X3-X6 in 2008-2010, which have typical
sampling of a few days over 2 yr, give $F_{\rm var}=10\pm 4$ \%. There
is no difference in $F_{\rm var}$ between the soft (0.5-2 keV) and
hard (2-7 keV) bands light curves.

Significant variations of the flux ratio (X-ray colour) are
observed, indicating changes in spectral shape. Since some
incompatibility in spectral calibration between XMM-Newton and Chandra
exists (see e.g., Lanzuisi et al 2013)\footnote{XMM-Newton slopes are
  generally found to be steeper by $\Delta\Gamma = $ 0.1-0.2 than
  Chandra for an object with $\Gamma\sim 2$ at the brightness of these
  sources.}, a direct comparison of the X-ray colours between the two
observatories is reserved. However, with XMM-Newton observations
alone, the X-ray colour variability is evident. Fig. 4a shows the
correlation diagram between the full-band (0.5-7 keV) flux and the
X-ray colour. There is no correlation between the two parameters but
the X-ray colours are not constant (even excluding the outlying
lowest-flux point, the constant hypothesis yields $\chi^2 =65.7$ for
31 degrees of freedom).

\subsection{Variability of absorption}


The observed X-ray colour variations could be caused partly by
variability in absorption as moderate absorption was detected in the
mean spectrum and the optical data taken 3 weeks apart showed a change
in reddening (Szokoly et al 2004). The XMM-Newton spectra taken from
the six intervals (Fig. 6) were studied to identify a spectral
parameter driving the X-ray colour variations. On examinig absorption
variability, the data were limited to the soft band (0.5-2 keV) to
avoid an uncertainty in the continuum modelling caused by the Fe K
emission and a possible spectral break in the hard band. An
absorbed power-law was fitted and photon index and absorbing column
density obtained for the six intervals are plotted in Fig 7. Note that
the first two intervals and the rest are separated by $\sim 6$ yr. The
hard band slope was also measured using the 2-7 keV data, including
the Fe K line modelled by a Gaussian but no absorption, as the degree
of absorption seen in the soft band does not affect the energies
greater than 2 keV.

Evidence for variability in absorption over the six intervals is found
to be weak ($\chi^2 = 6.5$ for 5 degrees of freedom). There is some
similarity seen in the temporal variation trends between $N_{\rm H}$
and the X-ray colour in Fig. 4 which suggests absorption may partly
drive the the X-ray colour variability. However, no drop in UV flux
was observed with the Optical Monitor (OM) of XMM-Newton in the X3
interval (Vagnetti et al, in prep.) where increased $N_{\rm H}$ is
suggested. This agrees with no variability in absorption, provided
same matter is responsible for both X-ray absorption and UV
extinction.

\subsection{Continuum slope variability}

The continuum slope should be the source of the X-ray colour changes,
if absorption is not a principal driver. The XMM-Newton spectra from
the six intervals however do not give any clear evidence for slope
changes, as shown in Fig. 7. The 0.5-2 keV and 2-7 keV slopes show
comparable values given the error bars of the measurements, in
agreement with the good fit of a single power-law for the mean
spectrum. Any spectral variability occurred in the 33 exposures might
be smeared out when they are binned into the six intervals, and the
spectral analysis cannot identify what caused the X-ray colour
variability. 

Although, as mentioned above, the error bars make it inconclusive,
when the soft- and hard-band slopes are plotted against flux in the
respective bands (Fig. 5b), a negative trend is seen for the soft-band
slopes. This is contrary to the``steeper when brighter'' trend often
seen in X-ray continuum slope of well-studied AGN.

\subsection{Fe K line variability}



\begin{figure}
\centerline{\includegraphics[width=0.45\textwidth,angle=0]{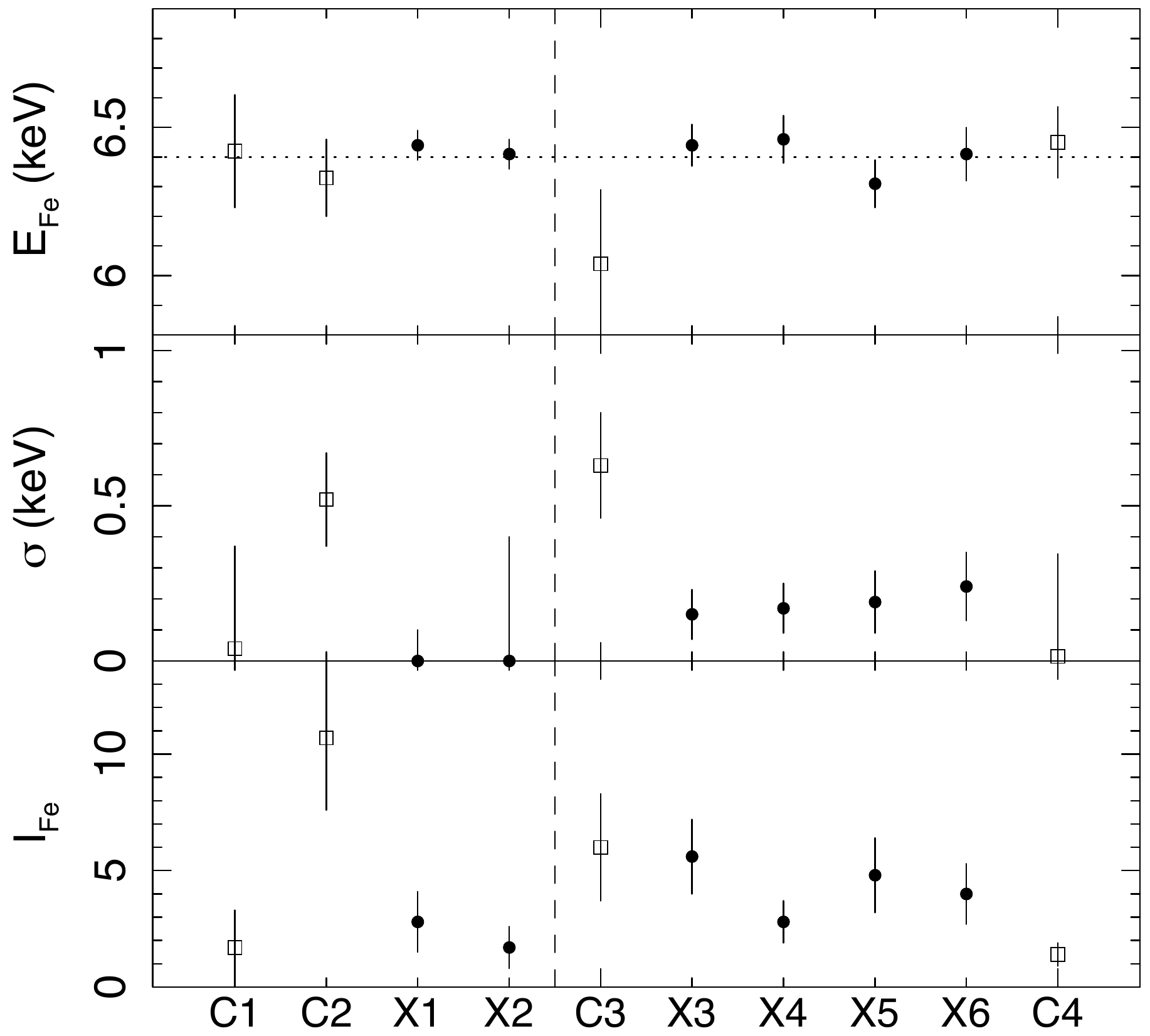}}
\caption{Measurements of the Fe K line parameters of PID 203: the
  rest-frame Gaussian centroid energy, Gaussian dispersion ($\sigma $)
  and flux, obtained from the six XMM (filled circles, see also
  Fig. 6) and four Chandra (open squares) intervals in time order. The
  dashed line indicates the 6 yr gap between the X2 and C3
  observations. Notable changes both in line shape and flux
  observed in the Chandra data are demonstrated in
  Fig. 9. The X-ray source flux dependence of these line parameters
  are shown in Fig. 10.}
\end{figure}


\begin{figure}
\centerline{\includegraphics[width=0.5\textwidth,angle=0]{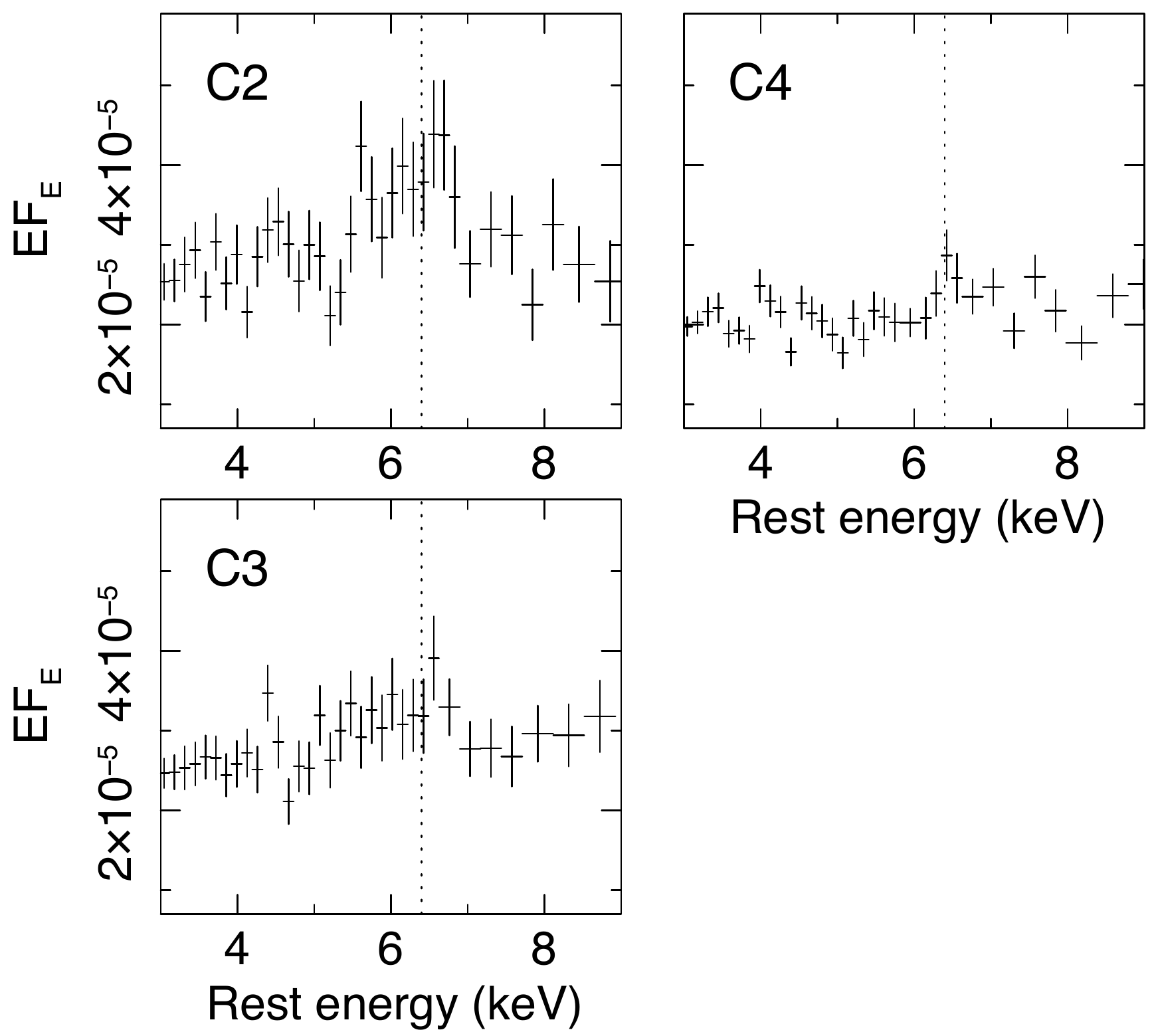}}
\caption{The rest-frame 3-9 keV spectra of PID 203 obtained from the
  Chandra ACIS observations during the C2, C3 and C4 intervals. The
  y-axis range is identical between the three plots. The dotted line
  indicates rest-frame 6.4 keV where a cold Fe K line is expected. The
  C2 spectrum shows apparently broad, strong Fe emission, the shape of
  which is reminiscent of that expected from a relativistic disk
  (e.g., Fabian et al 1989, see text for details), while the C4
  spectrum has a weak, narrow line. The C3 spectrum shows an Fe K
  feature weaker than that of C2 but also has a broad component as
  well as a narrow line. The line profile is strongly skewed redwards which
  causes a single Gaussian fit to give a centroid energy significantly
  lower than 6.4 keV (see Fig. 10). }
\end{figure}


\begin{figure}
\centerline{\includegraphics[width=0.37\textwidth,angle=0]{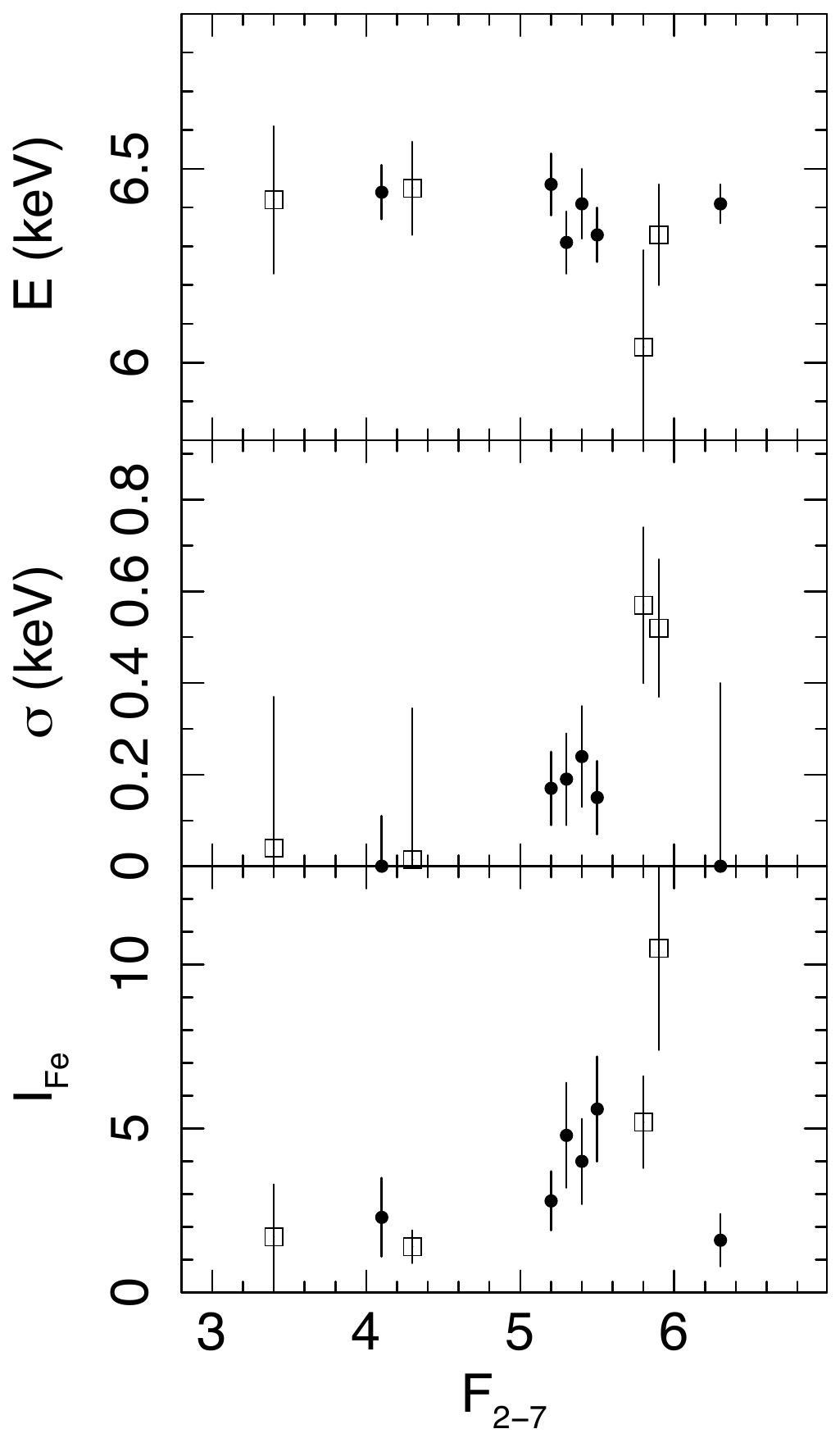}}
\caption{The Fe K line parameters of PID 203 measured in the six XMM
  (filled circles) and four Chandra (open squares) intervals, plotted
  against the 2-7 keV flux in units of $10^{-14}$ \ergpspsqcm. A
  significant ($>99$ \% confidence level) variability in line
  flux is observed. }
\end{figure}

The Fe K line variability was investigated using the six XMM intervals
as well as the four Chandra intervals. The Fe K feature was modelled
by a Gaussian and the continuum was determined by fitting a power-law
to the observed 2-6 keV data. Fig. 8 shows the rest-frame
line-centroid energy, the Gaussian dispersion, and the intensity in
time order. Among the XMM-Newton observations, the line emission is
found to be broader and stronger in the recent observations in
2008-2010 than in the two earlier observations in 2001-2002. More
dramatic changes appear to have occurred in Chandra observations: very
high flux acompanied with line broadening is observed in C2 compared
to the other intervals (Fig. 9). This strong, broad line disappeared
in the first XMM-Newton observation (X1) eight months later when the
Fe K returned to a narrow, weak line, albeit the data are noisy. 

The line parameters are plotted against the hard-band (2-7 keV) flux
(Fig. 10). A steep increase of line flux, rather than a continuous
trend, as increasing continuum flux in the range of (5-6)$\times
10^{-14}$ \ergpspsqcm\ in $F_{2-7}$ is observed. This rise in line
flux is linked with the line broadening and the shift of line centroid
to lower energies, indicating that enhancements of the red wing of the
line cause the flux increases. In the context of a relativistically
broadened line from the accretion disk, they are equivalent to
enhancements of relativisitic effects due to strong gravity around the
black hole. A simple interpretation would be that the X-ray source and
the Fe line emitting region move inwards where both disk emissivity
and gravitational effects are large as increasing continuum
flux. However, this simple picture breaks down when the data point of
the highest continuum flux (X2) is included. X2 is the observation in
2002 before the 6 yr gap, where the line is unresolved and weak (see
Fig. 6).

The strongest and broadest line feature observed in C2 was examined in
the context of the line from the relativistic disk. The {\tt diskline}
model for a Schwarzschild black hole (Fabian et al 1989) was used,
correcting the energy-scale for the redshift. The model for a
rest-frame 6.4 keV line with the line emitting radii of 6-50 $r_g$
(where $r_g = GM/c^2$), inclination of the disk of $30^{\circ}$ agrees
with the observed line shape (Fig. 9). No extreme broadening in a
rapidly spinning black hole is required. However, the very large EW of
the line (0.6 keV) is hard to explain if the innermost radius of the
line emitting region is as large as 6 $r_g$ (where an enhancement of
reflection due to light bending is diminished, Miniutti \& Fabian 2004).

\section{PID 319}


\begin{figure}
\centerline{\includegraphics[width=0.4\textwidth,angle=0]{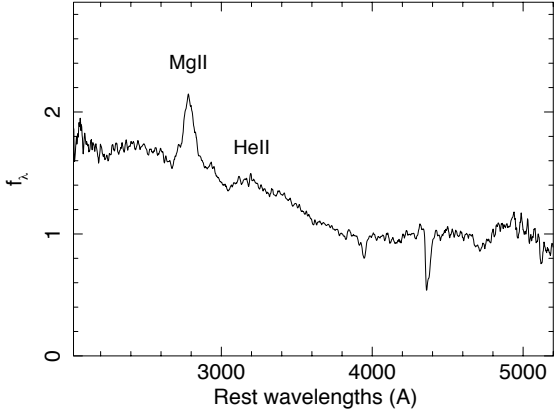}}
\caption{The UV/optical spectrum of PID319, taken at the VLT with the
  FORS spectrosgraphs (Szokoly et al 2004). The horizontal axis
  represents the rest-frame wavelengths in \AA, assuming the
  galaxy redshift of $z=0.742$. The flux density $f_{\lambda }$ is
  in units of $10^{-17}$ erg cm$^{-2}$ s$^{-1}$ \AA$^{-1}$.}
\end{figure}


\begin{figure}
\hbox{{\includegraphics[width=0.24\textwidth,angle=0]{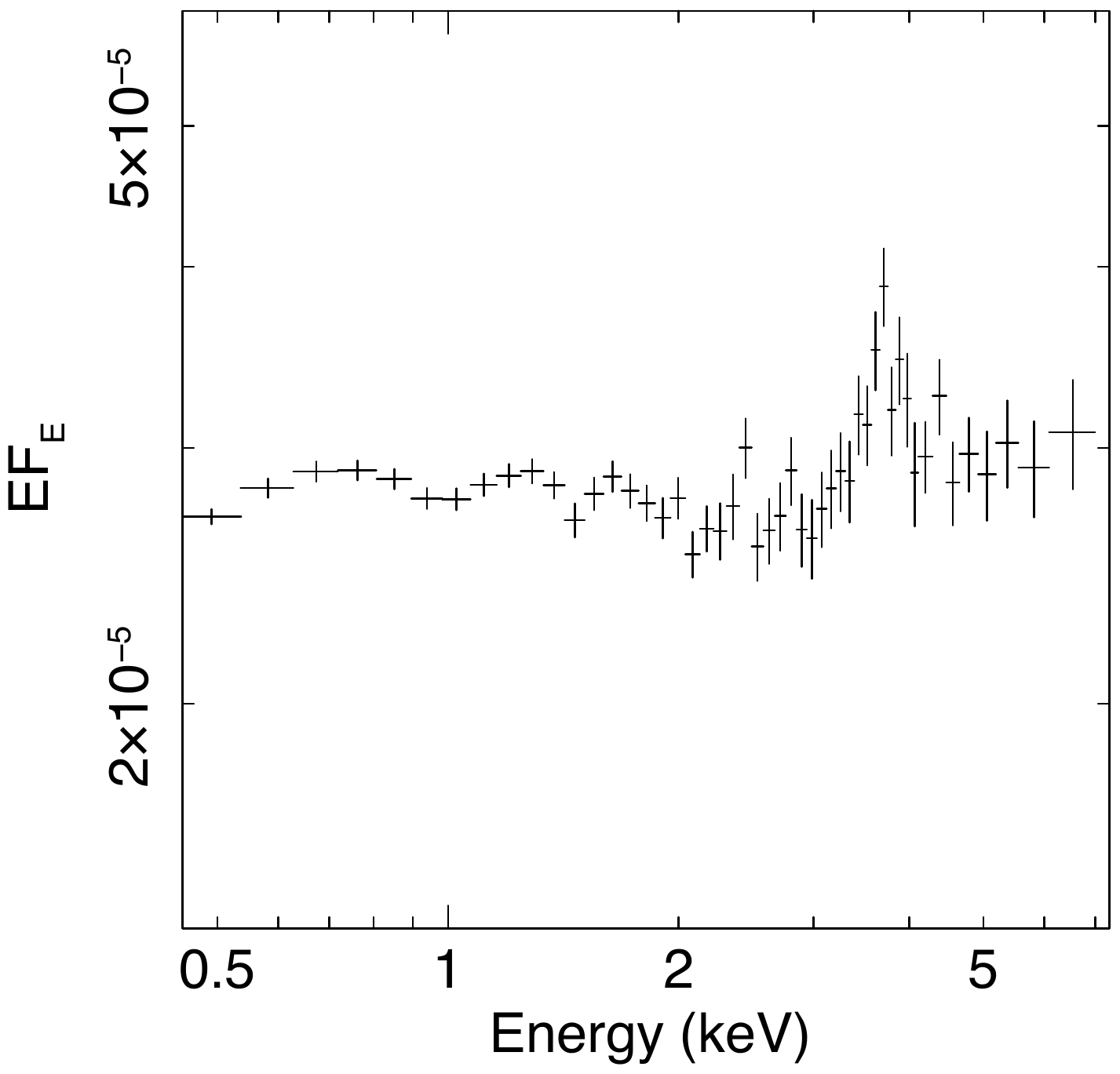}}
{\includegraphics[width=0.24\textwidth,angle=0]{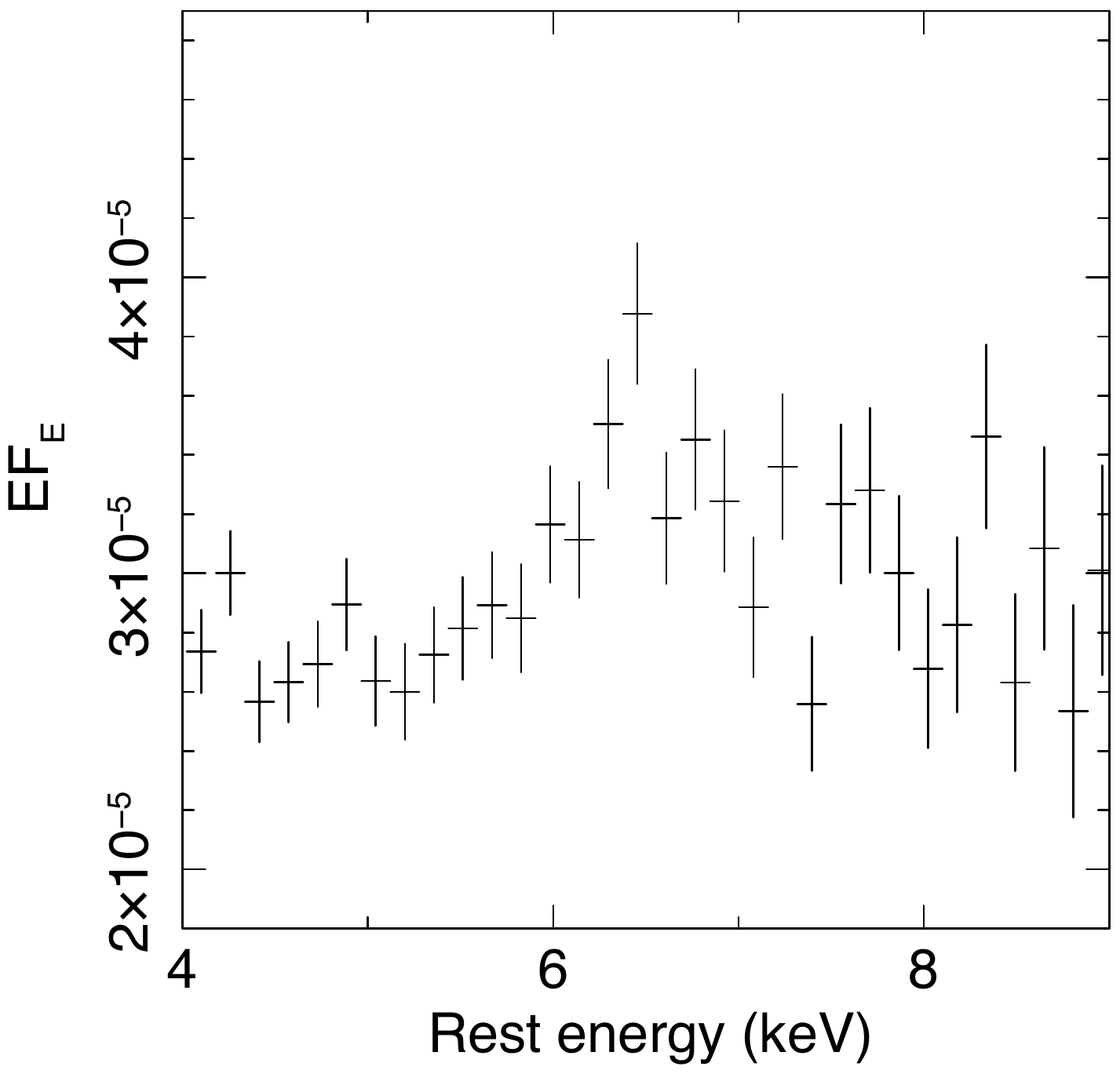}}}
\caption{Left: The 0.5-7 keV spectrum of PID 319, obtained from the
  full exposure of three EPIC cameras. The data are plotted in flux
  units of $EF_{E}$ as a function of energy as observed. The energy
  range corresponds to the rest-frame 0.87-12.2 keV. The small decline
  at the lowest energies is due to the Galactic absorption. The
  spectral flattening towards higher energies (see text) can be
  noticed.  Right: Details of the Fe K band spectrum plotted in the
  rest-frame 4-9 keV range. The Fe K line shows a moderately broad
  profile.}
\end{figure}


\begin{figure}
\hbox{{\includegraphics[width=0.237\textwidth,angle=0]{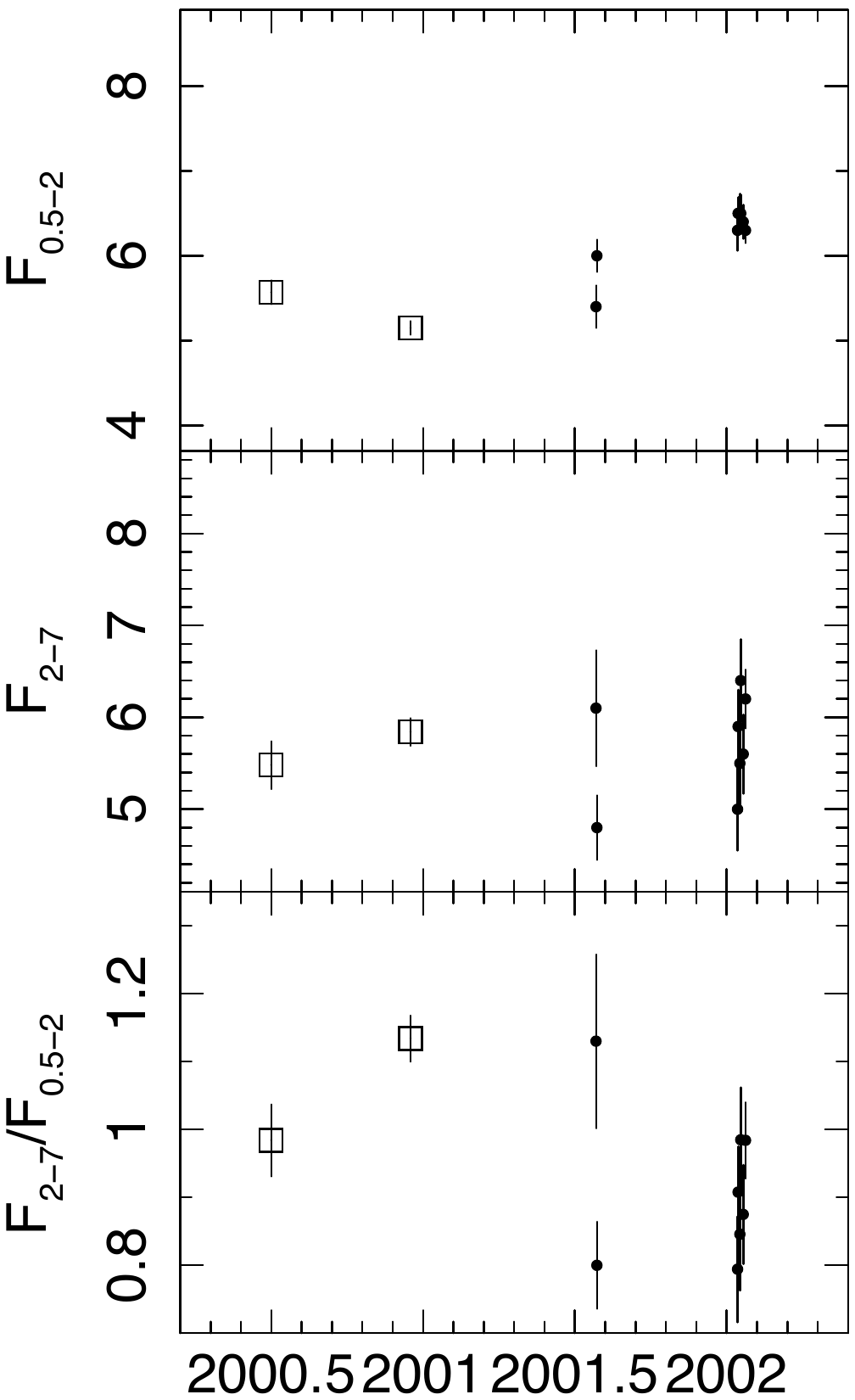}}{\includegraphics[width=0.25\textwidth,angle=0]{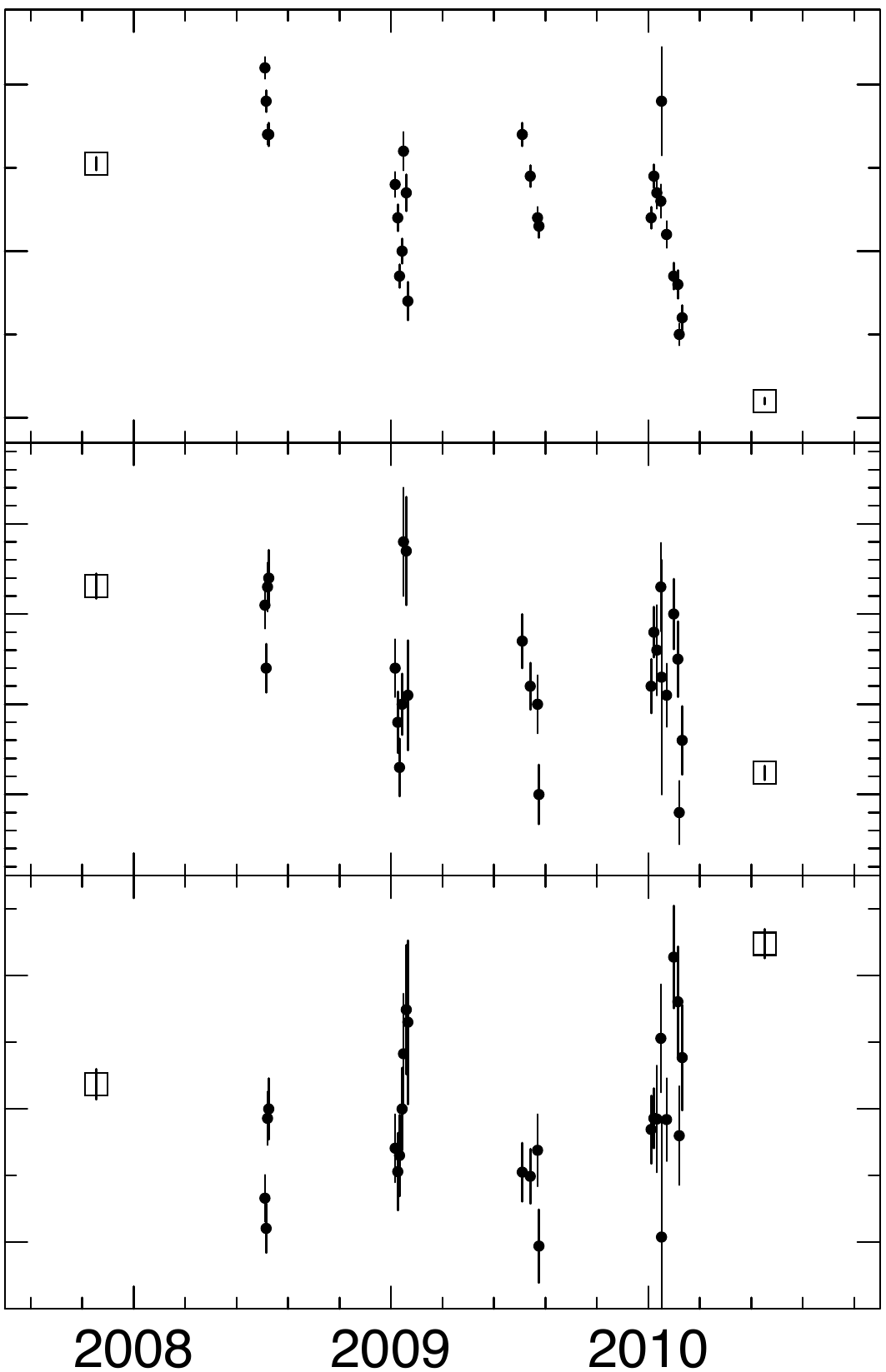}}}
\caption{The observed 0.5-2 keV and 2-7 keV flux and the flux ratio
  (2-7 keV/ 0.5-2 keV) variations of PID 319, obtained from the
  XMM-Newton (filled circles) and Chandra (open squares)
  observations. The time axis is in year. }
\end{figure}

\begin{figure}
\centerline{\includegraphics[width=0.37\textwidth,angle=0]{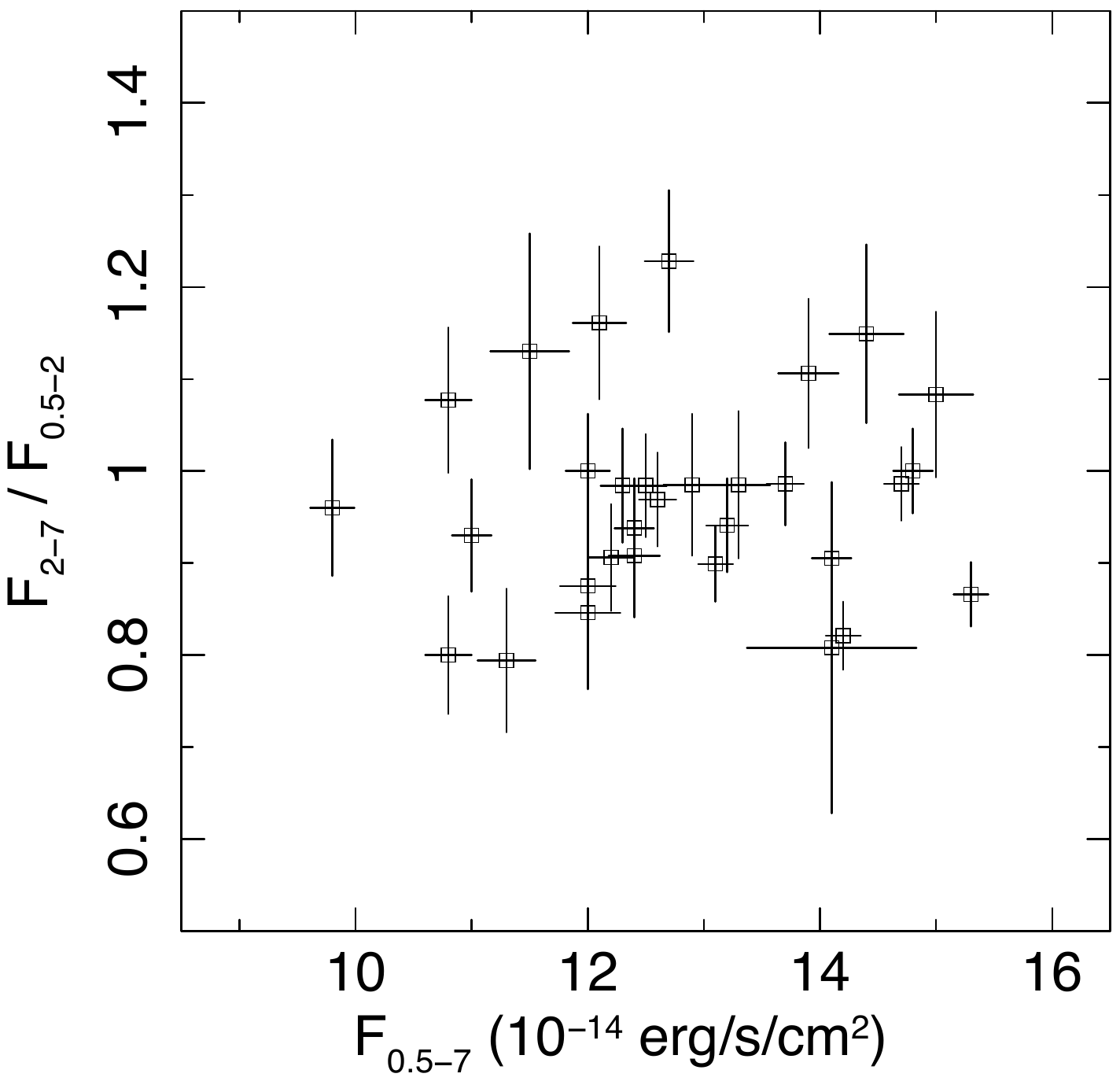}}
\caption{Plot of X-ray colour ($F_{\rm 2-7}/F_{\rm 0.5-2}$) against
  the 0.5-7 keV band flux for the 33 XMM exposures for PID 319. No
  correlation between the X-ray colour and the observed flux is seen.}
\end{figure}

PID 319 is the brightest X-ray source in the XMM-CDFS and is
identified with a QSO at redshift $z=0.742$. The optical (rest-frame
UV) spectrum presented in Szokoly et al (2004) shows clear, broad Mg
{\sc ii} emission (FWHM $\simeq 11,000$ km s$^{-1}$) on a blue
continuum, typical of broad-line QSOs.

The UV emission of PID 319 observed with the OM instrument (Antonucci
et al 2014) showed a notable decrease in luminoisty from $1\times
10^{30}$ erg~s$^{-1}$~Hz$^{-1}$ during the 2001-2002 period (X1-X2) to
$0.6\times 10^{30}$ erg~s$^{-1}$~Hz$^{-1}$ during the 2007-2010 period
(X3-X6) at 2500 \AA. The optical-to-X-ray spectral slope showed a
significant change correspondingly from $\alpha_{\rm OX}=-1.45\pm
0.01$ to $-1.33\pm 0.01$ (Vagnetti et al in prep.).

\subsection{The mean spectrum}


The mean X-ray spectrum of PID 319 obtained from the full exposure
(Fig. 12) shows no evidence for excess cold absorption ($N_{\rm
  H}<6\times 10^{19}$ \psqcm, 90 \% upper limit). The Fe K
feature (Fig. 12) is found at the rest-frame energy of $6.48\pm 0.07$
keV with significant broadening ($\sigma = 0.29^{+0.12}_{-0.09}$ keV,
or FWHM$\sim $21,000 km s$^{-1}$) when a Gaussian is fitted (Table
5). The line flux is estimated to be $4.2^{+1.1}_{-0.9}\times 10^{-7}$
\phpspsqcm, corresponding to a rest-frame EW of $0.20^{+0.05}_{-0.05}$
keV.

The continuum slope flattens toward higher energies. When a broken
power-law is fitted, the photon index changes from $\Gamma_1 =
2.09^{+0.01}_{-0.01}$ to $\Gamma_2 =1.88^{+0.03}_{-0.03}$ below and
above the rest-frame break energy of $E_{\rm br} =
3.74^{+0.41}_{-0.33}$ keV ($2.15^{+0.24}_{-0.19}$ keV in the observed
frame). This broken power-law model gives an improved fit ($\chi^2 =
515.3$ for 452 dof) relative to a single power-law (with $\Gamma =
2.05\pm 0.01$, $\chi^2 = 554.7$ for 454 dof) for the 0.5-7 keV
data. The steepning soft portion of the spectrum may be identified as
a {\it soft excess} that is often discussed for nearby AGN.

While this curved continuum may be an intrinsic feature of the X-ray
source, a plausible explanation is augmented emission in the hard band
(i.e., above the break energy) due to reflection from cold matter, as
expected from the presence of the Fe K line. Since, unlike PID 203,
PID 319 shows no evidence for absorption, which would introduce
uncertainty in slope measurements at lower energies, we examined the
reflection scenario in details, using the two similar but slightly
different reflection models. We modelled the 0.5-7 keV data with a sum
of an illuminating power-law and its reflection spectrum described by
{\tt reflionx} (Ross \& Fabian 2005) or {\tt xillver} (Garc\'ia et al
2013). Here no excess absorption is required and no relativistic
blurring of the reflection spectrum is applied. The ionization
parameter $\xi $ was left free in the spectral fits, but in both
cases, the reflecting matter was found to be cold ($\xi $ was pegged
to the lowest value 10 erg\thinspace s$^{-1}$\thinspace cm in {\tt
  reflionx} while $\xi = 5.8^{+3.6}_{-2.6}$ erg\thinspace
s$^{-1}$\thinspace cm in {\tt xillver}) to give good fits to the data
($\chi^2 = 534.5$ ({\tt reflionx}) and $\chi^2 = 531.2$ ({\tt
  xillver}) for 453 degrees of freedom). The photon index of the
illuminating power-law source was found to be $\Gamma = 2.06\pm 0.01$
({\tt reflionx}) or $\Gamma= 2.10\pm 0.01$ ({\tt xillver}). The
fraction of the 2-7 keV flux due to the reflected light was found to
be 10 or 14\%, in a good agreement with that is expected from the
albedo from a thick, cold matter subtending in $2\pi$ in the same
observed band ($\approx 10$ \%), means that the reflection
scenario is reasonable for explaining the spectral break.

\subsection{X-ray flux and colour variability}


The source flux measurements in the observed 0.5-2 keV ($F_{\rm
  0.5-2}$) and 2-7 keV ($F_{\rm 2-7}$) bands and the X-ray colour
($F_{\rm 2-7}/F_{\rm 0.5-2}$) for the 33 XMM exposures are plotted in
Fig.~13. As done for PID 203, fractional variability amplitudes was
derived for the XMM-Newton and Chandra combined 10 intervals ($F_{\rm
  var}=15\pm 5$ \%) and the 25 XMM-Newton exposures in the X3-X6
intervals ($F_{\rm var}=11\pm 4$ \%).  $F_{\rm var}$ of the soft and
hard bands are found to be comparable, like in PID 203.

The X-ray colour measured in the 33 XMM exposures is significantly
variable ($\chi^2 = 82.2$ for 32 degrees of freedom) but has no
correlation with the full band (0.5-7 keV) flux (Fig. 14).

\begin{figure}
\centerline{\includegraphics[width=0.37\textwidth,angle=0]{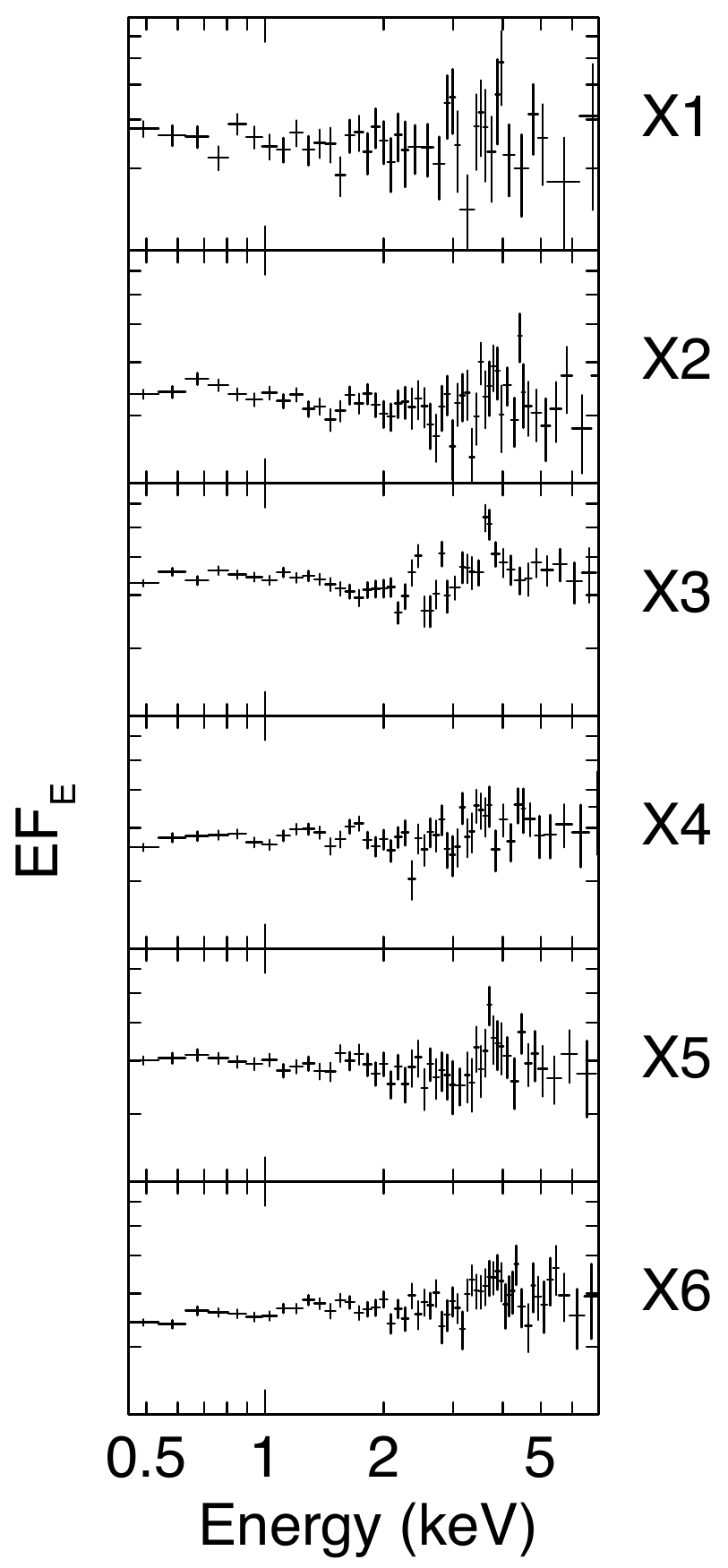}}
\caption{The XMM-Newton spectra of PID 319 observed in the six intervals. }
\end{figure}


\begin{figure}
\centerline{\includegraphics[width=0.4\textwidth,angle=0]{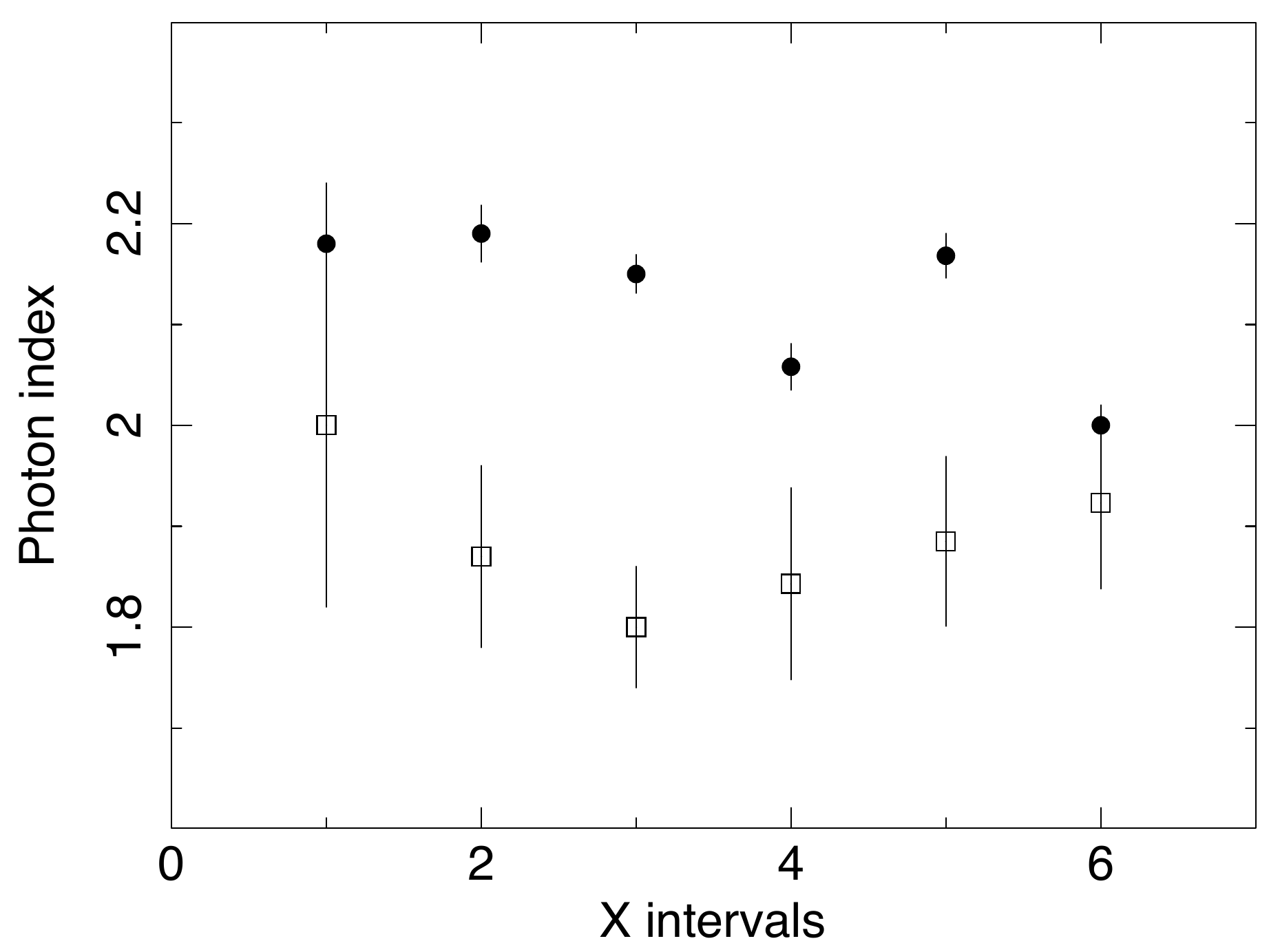}}
\smallskip
\centerline{\includegraphics[width=0.48\textwidth,angle=0]{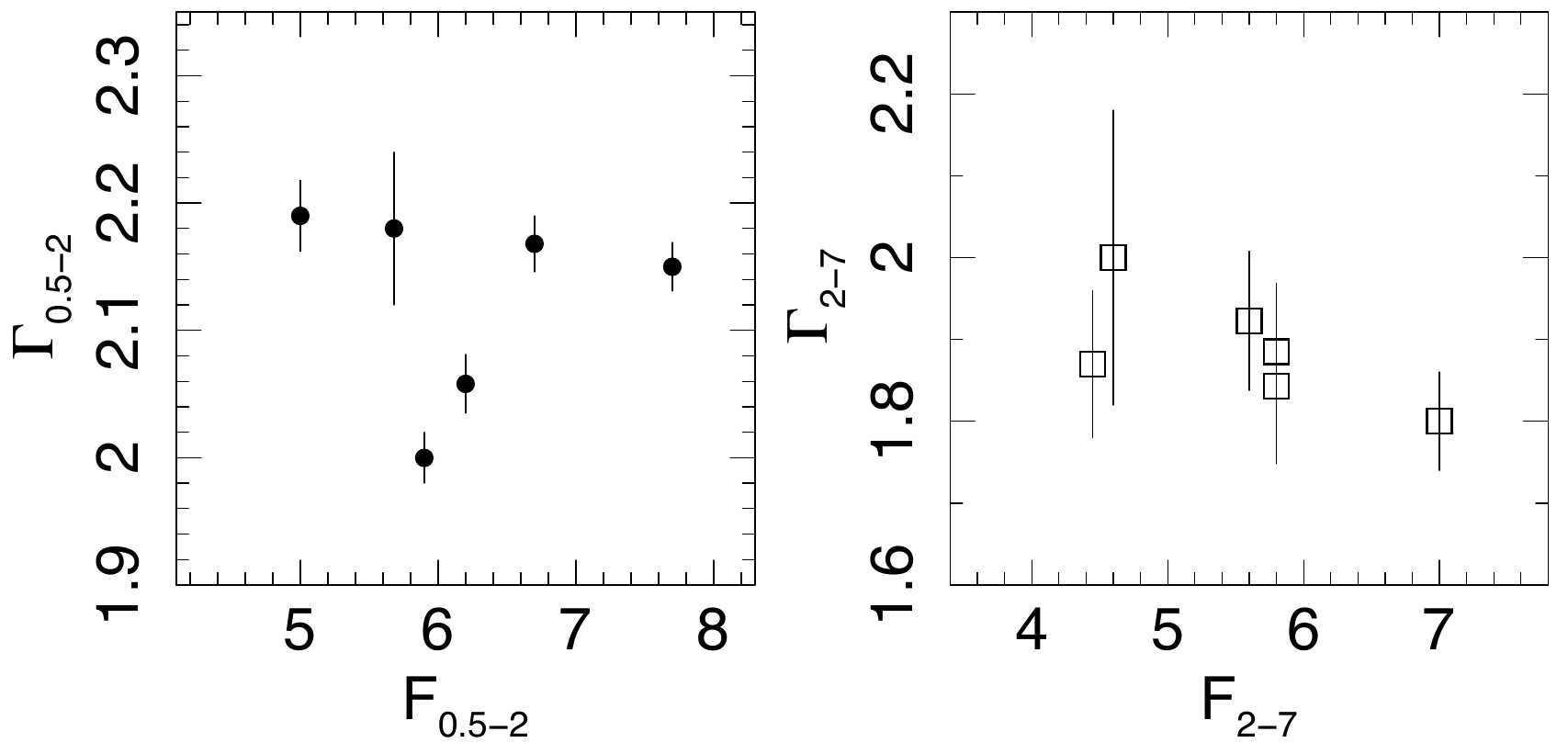}}
\caption{Upper panel: Photon index measurements for the six XMM
  intervals (X1-X6) in the 0.5-2 keV (filled circles) and the 2-7 keV
  (open squares) bands for PID 319. In X2-X5, the measured $\Gamma $
  differ significantly between the soft and hard bands, suggesting the
  presence of spectral break around 2 keV. Lower-left: Photon indices
  measured for the 0.5-2 keV continuum plotted against the measured
  flux in the same band for the six XMM intervals. Lower-right: Same
  as the left panel but in the 2-7 keV band. The fluxes are in units
  of $10^{-14}$ \ergpspsqcm. }
\end{figure}

\subsection{X-ray continuum shape}

We also investigated variability of the continuum slope within the
respective soft (0.5-2 keV) and hard (2-7 keV) bands, using the mean
spectra sampled at the six epochs (X1-X6, Fig. 15). As the spectral break
appears around observed 2 keV, we fit a power-law to the data below
and above 2 keV separately and recorded the photon indices
(Fig. 16). 

The soft- and hard-band slopes are not correlated and the hard-band
slope is always flatter than that in the soft band, suggesting that
the continuum slope pivots around 2 keV. Any correlation of the slope
with the flux is unclear, but, again, the ``steeper when brighter''
trend is not observed.

We investigated the spectra of X4 and X6 intervals, which show their
soft-band slopes clearly flatter than the others. No evidence
for absorption is found when an absorbed power-law is fitted, with the
90 \% upper limit of absorbing column density of $N_{\rm
  H}=1.7\times 10^{20}$ \psqcm. Therefore it is unlikely that the
spectral variability is caused by absorption.

\subsection{Fe K line variability}


\begin{figure}
\centerline{\includegraphics[width=0.45\textwidth,angle=0]{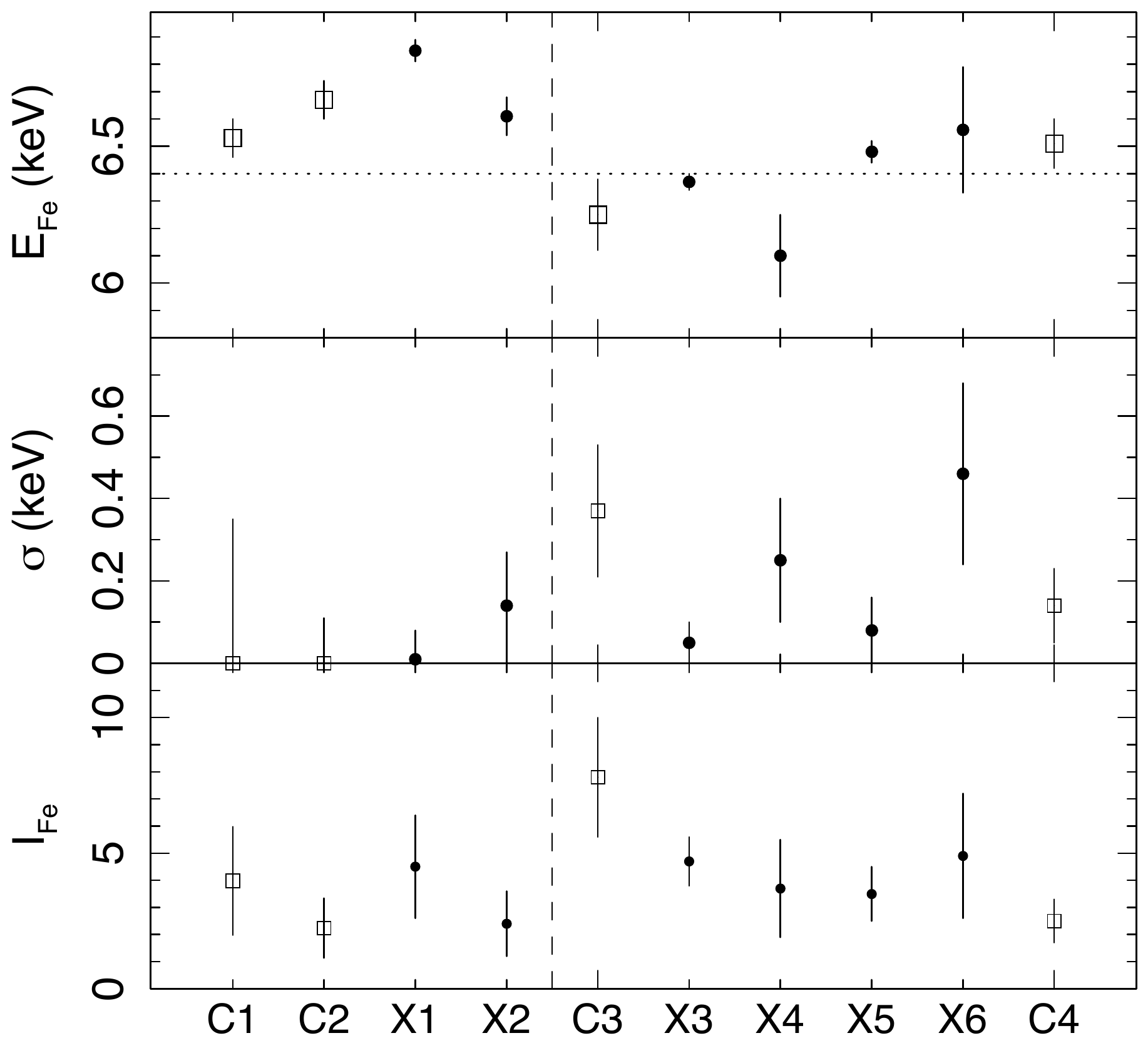}}
\caption{The Fe K line variability of PID 319, investigated for the
  four Chandra (open squares) and six XMM (filled circles)
  intervals. The emission line was modelled by a Gaussian, and the
  rest-frame line-centroid energy (upper panel), line width measured
  by the Gaussian dispersion (middle panel) and the normalization in
  unit of $10^{-7}$ photons s$^{-1}$ cm$^{-2}$. In the upper panel,
  the nominal line energy of 6.4 keV expected for cold Fe is indicated
  by dotted line. The dashed vertical line indicates the $\sim 6$ year
  gap between the 2000-2002 and 2007-2010 periods. }
\end{figure}

Variability of the Fe line was investigated using the spectra of the
six XMM and the four Chandra intervals in the same manner as for PID
203. The spectral data in the 2-6 keV band were modelled with a
power-law and a Gaussian.

Fitting results on the line parameters are shown in time order in
Fig. 17. Unresolved Fe lines with their centroid energy higher than
6.4 keV ($6.73\pm 0.15$ keV in weighted mean) are observed in the
2000-2002 period while broadened emission around 6.4 keV (weighted
mean of $6.41\pm 0.09$ keV) is consistently observed in the recent
2007-2010 intervals. This may imply that the physical condition of the
major Fe-line emitting region, the ionization condition, location
etc., changed during the 6-yr observation gap. The line width observed
in the full-exposure spectrum ($\sigma = 0.29$ keV, Fig. 12) is partly
due to these shifts in line energy. Within the six intervals in the
2007-2010 period, the line width may oscillate between the intervals
(i.e., approximately every half-yr), although the error bars are too
large to be conclusive.

On inspecting the line parameters as a function of 2-7 keV flux
(Fig. 18), the line centroid energy is not constant ($\chi^2 = $ 112
for 9 dof) and show flux dependence (the correlation coefficient is
$-0.70$) when the XMM-Newton and Chandra data combined together. A
constant hypothesis cannot be ruled out for the line width and flux,
given the large error bars. However, a positively correlated trend
between line and continuum flux can be seen. It is similar to the Fe
line of PID 203 that the line centroid is shifted to lower energies
when broadened.


\begin{figure}
\centerline{\includegraphics[width=0.37\textwidth,angle=0]{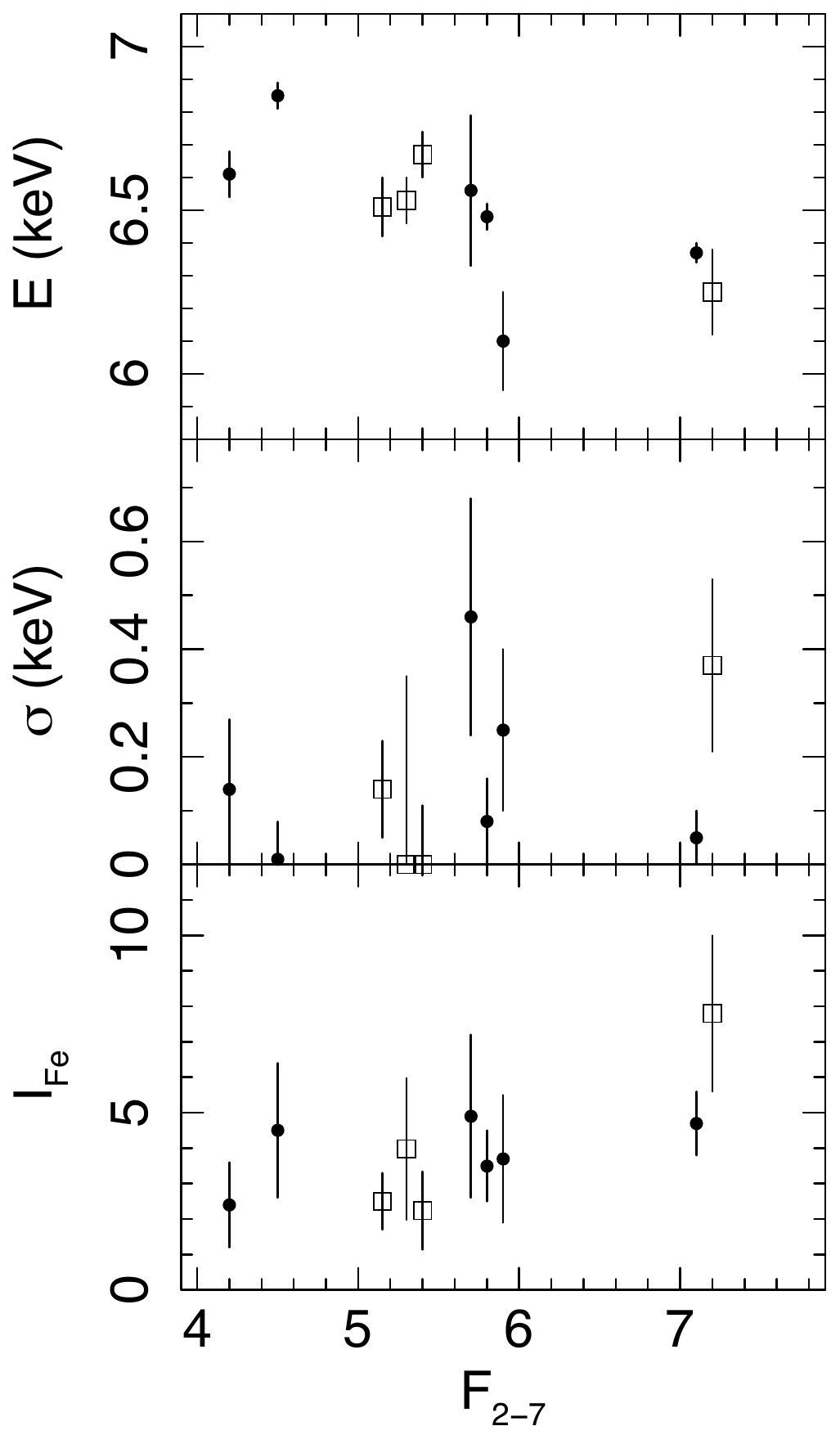}}
\caption{The Fe K line parameters of PID 319 measurd in the six XMM
  (filled circles) and four Chandra (open squares) intervals plotted
  against the 2-7 keV flux in units of $10^{-14}$ \ergpspsqcm. The
  $\chi^2$ values against a constant hypothesis are 112.0 for the line
  centroid, 10.1 for the line width, and 10.0 for the line flux,
  respectively, for 9 degrees of freedom (the XMM-Newton and Chandra
  datasets combined). }
\end{figure}

\section{Discussion}

The XMM-Newton long exposure of the CDF-S gave us an opportunity to
peek into the X-ray spectral/temporal properties of QSO with an X-ray
luminosity similar to those of low-z PG quasars beyond redshift $z\sim
0.5$ through the two bright X-ray sources PID 203 and PID 319 in the
field. Here we present a brief discussion of our fidings on X-ray
variability and the Fe K lines.

\subsection{X-ray variability}

Flux variability is a characteristic property of AGN and likely to
depend on the black hole mass ($M_{\rm BH}$). No reliable estimate of
$M_{\rm BH}$ are available for PID 203 and PID 319. The
Eddington-limit accretion places the lower limit of their $M_{\rm BH}$
of $10^7 M_{\odot}$, given their X-ray luminosity of $L_{\rm 2-10}\sim
10^{44}$ \ergps, assuming the typical bolometric correction. PG
quasars studied by Piconcelli et al (2005) with similar X-ray
luminoisties have typically $M_{\rm BH}\sim 10^8 M_{\odot}$. The
virial method cannot be applied for PID 203, as the optical
emission-line spectrum is severely altered by self-absorption. While
the Mg {\sc ii} width (FWHM $\simeq 1,1000$ km s$^{-1}$) can be used
for PID 319, the flux calibration in Szokoly et al (2004) data is
apparently problematic. A rough estimate of the continuum luminosity
using the R band magnitude (19.02 mag, Szokoly et al 2004) points to
that $M_{\rm BH}$ of PID 319 could be as large as $10^9 M_{\odot}$
(with the prescription of e.g., Shen et al 2011).

Temporal X-ray fluctuations in AGN generally exhibits red noise. This
means that $F_{\rm var}$ depends on time-scale of the measurements and
would be larger when sampled over a long time-scale. The two sources
show similar amplitudes of variability: $F_{\rm var}\simeq $15-17 \%
from the 10 XMM+Chandra intervals with half-yr sampling over 10 yr (net
4 yr) and $F_{\rm var}\simeq $10-11 \% from the 25 XMM-Newton
exposures over 2 yr (see Section 3.2 and 4.2). Within each XMM
interval where typical sampling is of a few days, $F_{\rm var}$ drops
to $\sim 8$ \% in both objects, in agreement with red noise. Among
many works on AGN X-ray variability, the monitoring programs of nearby
AGN with RXTE cover time-scales as long as 10 yr with dense sampling,
and are probably the best reference for our results. A comparison with
the results on various, bright AGN light curves obtained with RXTE in
Markowitz \& Edelson (2004) suggests that the $M_{\rm BH}$ of these
two CDFS sources may indeed be $10^8$-$10^9 M_{\odot}$. With similar
degree of X-ray variability and X-ray luminosities, the two CDFS AGN
are likely to have comparable $M_{\rm BH}$. A more
comprehensive X-ray variability study of CDFS sources, including these
two, will be presented in a separate paper (Paolillo et al. in prep.).

In comparison to nearby AGN, one notable property of X-ray variability
in these two QSOs is the lack of energy dependence in variability
amplitude. The light curves in the soft (0.5-2 keV) and hard (2-7 keV)
bands exhibit similar values of $F_{\rm var}$, contrary to that nearby
bright Seyfert galaxies often shows larger variability in the soft
band than in the hard band (Nandra et al 1997; Leighly 1999; Markowitz
\& Edelson 2001). This energy-dependence in variability found in
Seyfert galaxies is compatible with the ``steeper when brighter''
trend in the continuum slope often observed in those sources
(Markowitz \& Edelson 2001), in contrast to the two CDFS QSOs showing
neither the energy-dependent variability nor flux-dependent slope
changes. It is however interesting to note that the energy-dependence
of variability disappears on shorter time-scales, e.g., less than 40
ks for Seyfert galaxies, of which black hole masses are typically
$10^7 M_{\odot}$, studied by Ponti et al. (2012). If the black hole
masses of PID 203 and PID 319 are much larger than those Seyfert
galaxies, e.g., $\sim 10^9M_{\odot}$ as suspected, these QSOs might be
in the same regime, and a longer ($>10$ yr) monitoring may be needed to
see any deviation from the coherent varaibility in energy.

\begin{figure}
\hbox{\includegraphics[width=0.25\textwidth,angle=0]{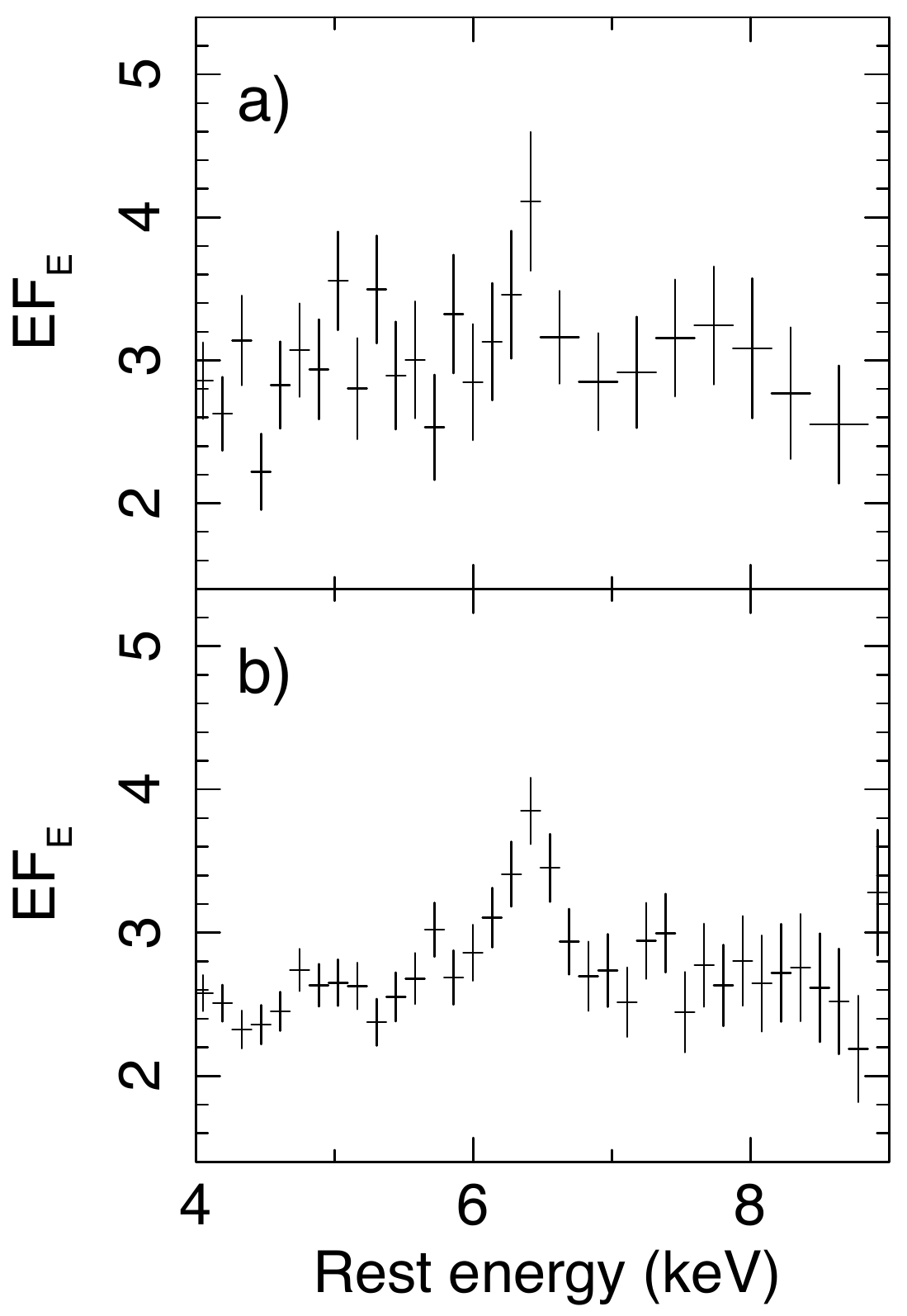}
\includegraphics[width=0.219\textwidth,angle=0]{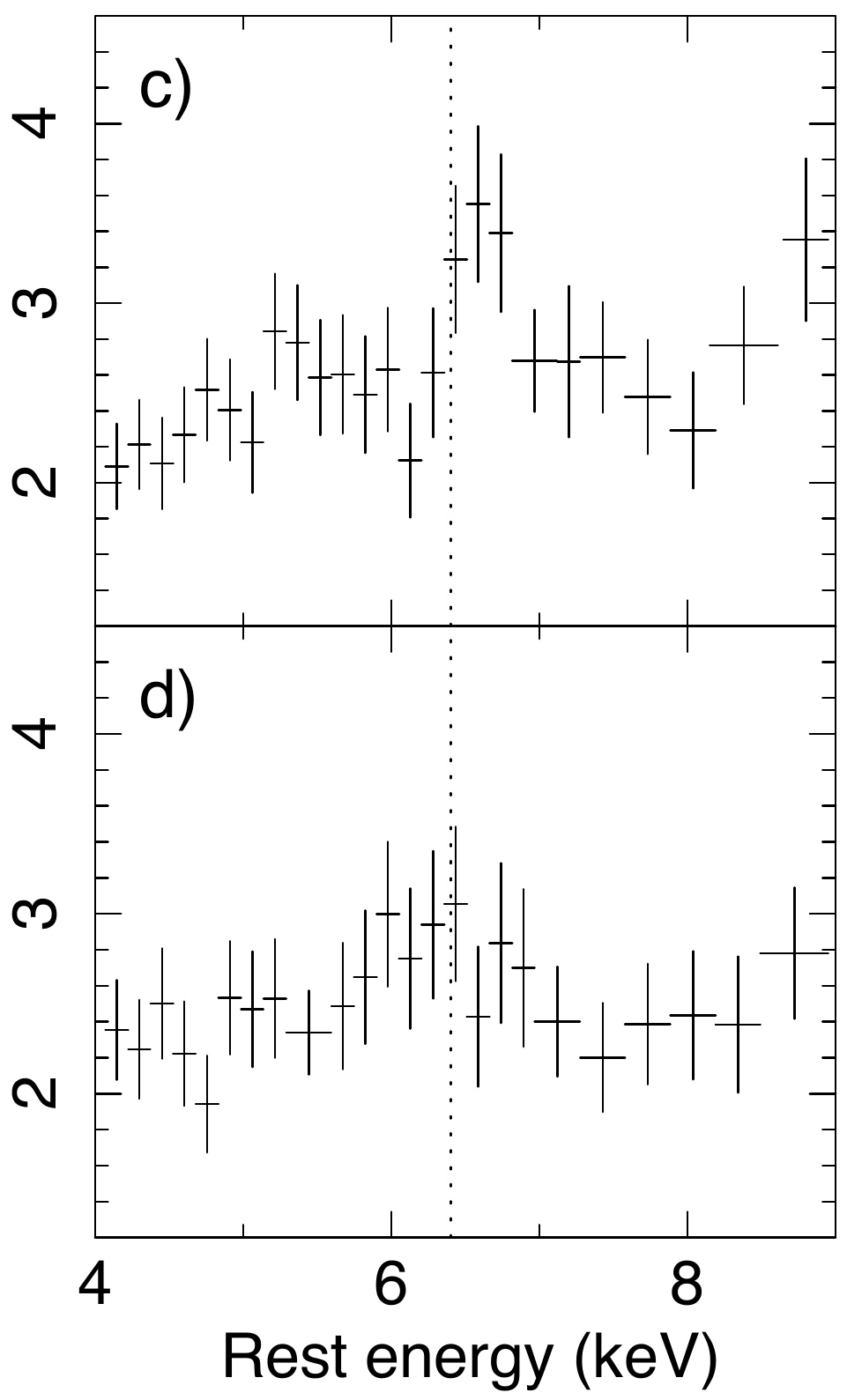}}
\caption{Examples of variable Fe K line profile on long and short
  time-scales, as observed with the same instrument. The left panel
  shows the Fe K band spectra of PID 203 observed with XMM-Newton in
  a) 2001-2002 (X1 and X2); and b) 2008-2010 (X3-X6) periods separated
  by 6 yr. As shown in Fig. 8, the line of a) is unresolved ($\sigma <
  0.12$ keV) while moderate but significant broadening ($\sigma =
  0.18\pm 0.04$ keV) is observed in b). The right panel shows the
  Chandra data of PID 319 during the first (c) and second (d) halves
  of the C4 interval with a 1 Ms exposure each. The C4 interval spans
  over 5 months (see Table 3), within which the line shape appears to
  change in a few months: the line centroid shifted from $6.56\pm
  0.05$ keV to $6.21\pm 0.17$ keV and the line profile broadened from
  $\sigma <0.14$ keV to $\sigma = 0.28^{+0.18}_{-0.13}$ keV while both
  continuum and line fluxes remain nearly identical with $f_{\rm
    2-7}=5.1\times 10^{-14} / 5.0\times 10^{-14}$ \ergpspsqcm, and
  $I_{\rm Fe}=(3.2\pm 1.0)\times 10^{-7} / (3.7\pm 1.5)\times 10^{-7}$
  \phpspsqcm, respectively.}
\end{figure}

\subsection{Fe K lines}

The Fe K lines are strongly detected in the mean spectra of PID 203
and PID 319, with EW of $\sim 0.2$ keV. Moderate broadening of $\sigma
$ of 0.2-0.3 keV is observed. According to the EW(Fe K) - $L_{\rm X}$
relation observed for nearby AGN (e.g., Iwasawa \& Taniguchi 1993),
the EW of a narrow Fe K originating from distant cold matter, e.g., a
torus, in an AGN with $L_{2-10}\sim 1\times 10^{44}$ \ergps is
expected to be about 0.04 keV (Bianchi et al 2007; Page et al 2004;
Fukazawa et al 2011; Ricci et al 2013). If this relation holds at $z=$
0.5-0.7, narrow Fe K emission of the torus origin would be a minor
component in the two QSOs and the bulk of their lines has to be
produced in the accretion disk at smaller distance, as the moderate
broadening of the line profiles suggests. However, the EW $\sim 0.2$
keV observed in the two objects are still remarkable, even though they
originate from the accretion disk. Systematic studies of broad Fe K
lines in nearby AGN found the typical value of broad-line EW to be
$\sim 0.1$ keV (Nandra et al 2007; de la Calle et al 2010). The
results of stacking analysis for distant AGN ($z\sim 1$; Corral et al
2008; Chaudhary et al 2012; Falocco et al 2012) also give similar
values. High Fe abundance can boost the line strength but it seems
unlikely that these two bright QSOs in the CDFS happen to be both AGN
of exceptional metalicity. Whether they are somehow exceptionally
strong Fe K emitters or there is an evolution in Fe K line strength of
AGN will need more data at high redshifts.

When the Chandra data are combined with the XMM-Newton data, the Fe K
lines seem to be variable: the line flux in PID 203 and the line
centroid in PID 319 (Fig. 10 and Fig. 18). Even limiting to the data
from a single observatory, for example, the line width of PID 203
measured with XMM-Newton, which was unresolved in the eariler X1-X2
periods, became broader in X3-X6 intervals six years later, as shown
in Fig. 19a,b. The C4 interval is of the last 2 Ms observation with
Chandra. When this interval is divided into two parts with a 1-Ms
exposure each, which still provides a comparable or better quality
spectrum that the other Chandra intervals, the line of PID 319 appears
to change its shape in three months (Fig. 19c,d) while the continuun
flux changes little. If these pieces of marginal evidence were 
to be real, they would support the accretion disk origin for the Fe K
lines. Given the moderate broadening, the primary line-emitting region
is not strongly relativistic but at radii small enough to receive
strong radiation from the central source. The dynamical time-scale of
a Keplerian disk around a $10^9 M_{\odot}$ black hole at 40 $r_g$ is
$T_{\rm d}\sim 3(R/40r_{\rm g})^{3/2}(M/10^9 M_{\odot})$ months, where
$R$ is the distance from the black hole with a mass of $M$. Thus, Fe K
line variability may be naturally expected if the line emission is
produced in the accretion disk at a few tens $r_g$.

While the Fe line energy remains at 6.4 keV in PID 203, the line in
PID 319 appears to originate from highy ionized matter in some
occasions, especially in 2000-2002. The four observations with
XMM-Newton and Chandra observations in that period showed consistently
that the line was found at 6.5-6.9 keV and unresolved. Evidence that
highly ionized Fe K (Fe {\sc xxv}, Fe {\sc xxvi}) may be formed in AGN
with a high Eddington ratio has been reported (e.g., Porquet et
al. 2004 and Inoue et al. 2007 for PG quasars; Iwasawa et al 2012 for
COSMOS AGN). In this respect, the steeper $\alpha_{\rm OX}$ measured
for PID 319 ($\alpha_{\rm OX}\sim -1.4$, Section 4) than that for PID
203 ($\alpha_{\rm OX}\sim- 1.2$, Section 3), as measured by the OM
instrument (Vagnetti et al in prep.), may be indicative. Using the
correlated trend between the Eddington ratio ($\lambda_{\rm Edd}$) and
$\alpha_{\rm OX}$ (e.g., Lusso et al 2010 for XMM-COSMOS Type I AGN),
$\lambda_{\rm Edd}\sim 0.1$ can be deduced for PID 319 with the
bolometric correction of $k_{\rm bol}\simeq 30$ for the 2-10 keV
luminoisty. A black hole mass of $M_{\rm BH}\sim 5\times
10^8M_{\odot}$ would realize this $\lambda_{\rm Edd}$ when the above
bolometric correction is used for the mean 2-10 keV luminosity of PID
319 ($1.8\times 10^{44}$ \ergps). This Eddington ratio is however
marginally low in comparison to those with high-ionization Fe K in the
previous study.  The source emitted at higher UV luminosity and had a
steeper $\alpha_{\rm OX}$ in the 2001-2002 period, where the
high-ionization Fe K was evident, than in 2007-2010 (see Section
4). While the UV-strong state in 2001-2002 perhaps marks a
high-ionization condition in PID 319, whether the UV luminosity larger
by $\sim 65$ \% is sufficient to account for the change in the
ionization condition remains uncertain.

\begin{acknowledgements}
  This research made use of the data obtained from XMM-Newton and the
  Chandra X-ray Observatory. Matteo Guainazzi is thanked for providing
  us with the information on the XMM-Newton calibration status. Kenta
  Matsuoka and Tohru Nagao are thanked for useful discussion on the
  optical data. KI acknowledge support by DGI of the Spanish
  Ministerio de Econom\'ia y Competitividad (MINECO) under grant
  AYA2013-47447-C3-2-P. WNB thanks the NASA ADP grant NNX10AC99G. We
  acknowledge financial contribution from the agreement ASI-INAF
  I/009/10/0.
\end{acknowledgements}

\end{document}